\documentclass[twocolumn]{aastex63}

\newcommand\teff{T_\mathrm{eff}}
\newcommand\logg{\log{g}}
\newcommand\vt{v_{t}}

\newcommand{\vinicrit}{v_\mathrm{ini}/v_\mathrm{crit}}

\received{18 May 2021}
\revised{01 July 2021}
\accepted{09 July 2021}

\shorttitle{Chemical abundances of the Phoenix stellar stream}
\shortauthors{Casey et al.}

\begin{document}

\title{Signature of a massive rotating metal-poor star imprinted in the Phoenix stellar stream\footnote{This paper includes data gathered with the 6.5 meter Magellan Telescopes located at Las Campanas Observatory, Chile.}}

\correspondingauthor{Andrew R. Casey}
\email{andrew.casey@monash.edu}

\author[0000-0003-0174-0564]{Andrew~R.~Casey}
\affiliation{Center of Excellence for Astrophysics in Three Dimensions (ASTRO-3D),
			 Australia}
\affiliation{School of Physics \& Astronomy, Monash University, 
			 Wellington Rd, Clayton 3800, Victoria, Australia}
			 
\author[0000-0002-4863-8842]{Alexander~P.~Ji}
\affiliation{Observatories of the Carnegie Institution for Science, 
			 813 Santa Barbara St., Pasadena, CA 91101, USA}

\author[0000-0001-6154-8983]{Terese~T.~Hansen}
\affiliation{George P. and Cynthia Woods Mitchell Institute for Fundamental Physics and Astronomy, 
			 and Department of Physics and Astronomy, 
			 Texas A\&M University, College Station, TX, 77843, USA}

\author[0000-0002-9110-6163]{Ting~S.~Li}
\affiliation{Observatories of the Carnegie Institution for Science, 
			 813 Santa Barbara St., Pasadena, CA 91101, USA}
\affiliation{Department of Astrophysical Sciences, Princeton University, 
			 Princeton, NJ 08544, USA}
\affiliation{NHFP Einstein Fellow}			 

\author[0000-0003-2644-135X]{Sergey~E.~Koposov}
\affiliation{Institute for Astronomy, University of Edinburgh, Royal Observatory, Blackford Hill, Edinburgh EH9 3HJ, UK}			 
\affiliation{Institute of Astronomy, University of Cambridge, 
			 Madingley Road, Cambridge CB3 0HA, UK}
\affiliation{McWilliams Center for Cosmology, Carnegie Mellon University, 
			 5000 Forbes Ave, Pittsburgh, PA 15213, USA}

\author[0000-0001-7019-649X]{Gary~S.~Da~Costa}
\affiliation{Center of Excellence for Astrophysics in Three Dimensions (ASTRO-3D),
			 Australia}
\affiliation{Research School of Astronomy and Astrophysics, 
			 Australian National University, Canberra, ACT 0200, Australia}

\author[0000-0001-7516-4016]{Joss~Bland-Hawthorn}
\affiliation{Center of Excellence for Astrophysics in Three Dimensions (ASTRO-3D),
			 Australia}
\affiliation{Sydney Institute for Astronomy, School of Physics, A28, 
			 The University of Sydney, NSW 2006, Australia}

\author[0000-0001-8536-0547]{Lara~Cullinane}
\affiliation{Research School of Astronomy and Astrophysics, 
			 Australian National University, Canberra, ACT 0200, Australia}
			 
\author[0000-0002-8448-5505]{Denis~Erkal}			 
\affiliation{Department of Physics, University of Surrey, Guildford GU2 7XH, UK}

\author[0000-0003-3081-9319]{Geraint~F.~Lewis}
\affiliation{Sydney Institute for Astronomy, School of Physics, A28, 
			 The University of Sydney, NSW 2006, Australia}

\author[0000-0003-0120-0808]{Kyler~Kuehn}			 
\affiliation{Lowell Observatory, 1400 W Mars Hill Rd, Flagstaff,  AZ 86001, USA}
\affiliation{Australian Astronomical Optics, Faculty of Science and Engineering, Macquarie University, Macquarie Park, NSW 2113, Australia}

\author[0000-0002-6529-8093]{Dougal~Mackey}
\affiliation{Center of Excellence for Astrophysics in Three Dimensions (ASTRO-3D),
			 Australia}
\affiliation{Research School of Astronomy and Astrophysics, 
			 Australian National University, Canberra, ACT 0200, Australia}

\author[0000-0002-3430-4163]{Sarah~L.~Martell}
\affiliation{Center of Excellence for Astrophysics in Three Dimensions (ASTRO-3D),
			 Australia}
\affiliation{School of Physics, 
			 University of New South Wales, Sydney, NSW 2052, Australia}

\author[0000-0002-6021-8760]{Andrew~B.~Pace}
\affiliation{McWilliams Center for Cosmology, 
			 Carnegie Mellon University, 5000 Forbes Ave, Pittsburgh, PA 15213, USA}
			 
\author[0000-0002-8165-2507]{Jeffrey~D.~Simpson}
\affiliation{Center of Excellence for Astrophysics in Three Dimensions (ASTRO-3D),
			 Australia}
\affiliation{School of Physics, 
			 University of New South Wales, Sydney, NSW 2052, Australia}

\author[0000-0003-1124-8477]{Daniel~B.~Zucker}
\affiliation{Center of Excellence for Astrophysics in Three Dimensions (ASTRO-3D),
			 Australia}
\affiliation{Department of Physics \& Astronomy, 
			 Macquarie University, Sydney, NSW 2109, Australia}
\affiliation{Macquarie University Research Centre for Astronomy, 
			 Astrophysics \& Astrophotonics, Sydney, NSW 2109, Australia}
			 



\newcommand\meanFeIIH{[\mathrm{Fe\,II/H}] = -2.71 \pm 0.07}
\newcommand{\spread}[5]{\sigma(\mathrm{[#1~#2/H]}) = #3^{#4}_{#5}}
\newcommand\spreadAlIH{\spread{Al}{I}{0.06}{+0.23}{-0.04}}
\newcommand\spreadFeIIH{\spread{Fe}{II}{0.04}{+0.11}{-0.03}}
\newcommand\spreadBaIIH{\spread{Ba}{II}{0.03}{+0.10}{-0.02}}
\newcommand\spreadMnIH{\spread{Mn}{I}{0.05}{+0.35}{-0.04}}
\newcommand\spreadCaIH{\spread{Ca}{I}{0.13}{+0.12}{-0.06}}
\newcommand\spreadSrIIH{\spread{Sr}{II}{0.38}{+0.33}{-0.21}}

\begin{abstract}
The Phoenix stellar stream has a low intrinsic dispersion in velocity and metallicity that implies the progenitor was probably a low mass globular cluster. In this work we use Magellan/MIKE high-dispersion spectroscopy of eight Phoenix stream red giants to confirm this scenario. In particular, we find negligible intrinsic scatter in metallicity ($\spreadFeIIH$) and a large peak-to-peak range in [Na/Fe] and [Al/Fe] abundance ratios, consistent with the light element abundance patterns seen in the most metal-poor globular clusters. However, unlike any other globular cluster, we also find an intrinsic spread in [Sr\,II/Fe] spanning $\sim$1~dex, while [Ba\,II/Fe] shows nearly no intrinsic spread ($\spreadBaIIH$). This abundance signature is best interpreted as slow neutron capture element production from a massive fast-rotating metal-poor star ($15-20\,\mathrm{M}_\odot$, $\vinicrit = 0.4$, $[\mathrm{Fe/H}] = -3.8$). The low inferred cluster mass suggests the system would have been unable to retain supernovae ejecta, implying that any massive fast-rotating metal-poor star that enriched the interstellar medium must have formed and evolved before the globular cluster formed. Neutron capture element production from asymptotic giant branch stars or magneto-rotational instabilities in core-collapse supernovae provide poor fits to the observations. We also report one Phoenix stream star to be a lithium-rich giant ($A(\mathrm{Li}) = 3.1 \pm 0.1$). At $[\mathrm{Fe/H}] = -2.93$ it is among the most metal-poor lithium-rich giants known.
\end{abstract}


\section{Introduction} 
\label{sec:introduction}

The Milky Way halo is littered with stars that have become gravitationally unbound from their host star cluster \citep[e.g.,][]{Ibata:1994,Belokurov:2007,Bonaca:2012}. This accretion process produces streams of stars in the Milky Way that lead and trail the progenitor. The Phoenix stellar stream is one recently discovered example, found with the Dark Energy Survey (DES) first data release \citep{Balbinot:2016}. The Phoenix stream is at a mean heliocentric distance of 19.1\,kpc \citep{Shipp:2019} and spans approximately 8$^\circ$ on the sky \citep{Balbinot:2016}. The relatively long arc and on-sky narrow width (54\,pc) implies the progenitor system was low-mass ($\approx{}3\times{}10^4\,M_\odot$) and had a small velocity dispersion \citep[i.e., dynamically cold,][]{Erkal:2016,Shipp:2019,Wan:2020}.

These kinematic features are consistent with the Phoenix stream being the tidally disrupted remains of a globular cluster. Low-resolution spectra of Phoenix stream members revealed a metallicity dispersion that is consistent with zero \citep{Li:2019,Wan:2020}, confirming a globular cluster origin. However, the mean metallicity of the Phoenix stream is remarkably low ($\mathrm{[Fe/H]}~=~-2.7$), making it $\sim\,$0.3\,dex more metal-poor than all known surviving globular clusters \citep{Wan:2020}. The detailed chemical abundances of stars in an ancient low-mass star cluster can be informative about nucleosynthetic events in the early universe. For this reason, the Phoenix stream represents a unique opportunity to study the formation of ancient globular clusters that are very low mass, and very metal-poor.

In this work we describe high-resolution spectroscopic observations of eight Phoenix stream members and present their detailed chemical abundances. In Section~\ref{sec:observations} we describe the observations, which were performed as part of the Southern Stellar Stream Spectroscopic Survey \citep[$S^5$;][]{Li:2019}. In Section~\ref{sec:methods} we summarise our methods \citep[which are expanded upon in ][]{Ji:2020} and our results are provided in Section~\ref{sec:results}. We discuss those results in Section~\ref{sec:discussion} and provide concluding statements in Section~\ref{sec:conclusions}.

\section{Observations}
\label{sec:observations}

We selected candidate Phoenix stream members based on whether their sky positions and proper motions were consistent with the orbit of the Phoenix stream \citep{Balbinot:2016,Gaia:2018,Shipp:2019,Li:2020}. We then restricted the candidates to those with $g-r$ colours \citep{Morganson:2018} and apparent magnitudes (at the distance of the stream) that were consistent with a 12\,Gyr very metal-poor isochrone \citep{Dotter:2008}. We acquired medium-resolution near-infrared spectra at the \ion{Ca}{2} triplet of candidate members using the AAOmega spectrograph on the 3.9\,m Anglo-Australian Telescope, of which 25 were consistent with stream membership based on their velocities and metallicities \citep[for further details see][]{Wan:2020}. 

We observed the brightest 8 Phoenix stream members between 2018-09-30 and 2019-10-19 with the MIKE spectrograph \citep{Bernstein:2003} on the 6.5\,m Magellan Clay telescope at Las Campanas Observatory, Chile. All observations were conducted in good conditions at low airmass ($<$\,1.25), with exposure times set to achieve a signal-to-noise (S/N) ratio of 20 per pixel at 400\,nm (Table~\ref{tab:observations}). We used the 1.0\arcsec\ slit for Phoenix 6 to suit poorer seeing conditions at the time, and the 0.7\arcsec\ wide slit for all other Phoenix observations.  We used 2 by 2 on-chip binning and slow read-out speed for all observations to reduce read noise. Biases and flat frames were taken in the afternoon, and arc frames were taken throughout the night. We used the \texttt{CarPy} reduction pipeline \citep{Kelson:2003} to reduce the data.


\begin{deluxetable*}{lccccccccc}
\centerwidetable
\tablecaption{Sky positions and exposure times for all stars observed. Designations from \citet{Wan:2020}. S/N has units pixel$^{-1}$.\label{tab:observations}}
\tablewidth{0pt}
\tablehead{
	\colhead{Star} &
	\colhead{R. A.} &
	\colhead{Dec.} &
	\colhead{Gaia DR2 Designation} &
	\colhead{Observed} &
	\colhead{Exp.} &
	\colhead{$g$} & 
	\colhead{Slit} & 
	\colhead{S/N} &
	\colhead{S/N} \\
	& \colhead{(hh:mm:ss)} & \colhead{(hh:mm:ss)} && Date & \colhead{(min)} & \colhead{(mag)} & \colhead{(arcsec)} & \colhead{450\,nm} & \colhead{650\,nm}
}
\startdata
Phoenix  1 & 01:23:48.36 & $-$53:57:27.4 & 4914426859986001920 & 2018-10-01 &  50 & 16.98 & 0.7 & 22 & 38 \\
Phoenix  2 & 01:24:36.27 & $-$53:40:01.2 & 4914446067079706624 & 2019-10-19 & 120 & 17.65 & 0.7 & 12 & 20 \\
Phoenix  3 & 01:25:55.15 & $-$53:17:35.1 & 4914527911976567424 & 2019-07-25 & 120 & 17.57 & 0.7 & 20 & 39 \\
Phoenix  6 & 01:39:20.84 & $-$49:09:11.7 & 4917862490225433984 & 2018-09-30 &  90 & 15.96 & 1.0 & 14 & 28 \\
Phoenix  7 & 01:42:44.22 & $-$47:29:05.2 & 4954034292475361280 & 2018-10-01 &  30 & 16.28 & 0.7 & 25 & 44 \\
Phoenix  8 & 01:41:53.37 & $-$47:06:51.6 & 4954245123830234240 & 2019-07-27 & 120 & 17.71 & 0.7 & 20 & 37 \\
Phoenix  9 & 01:48:16.06 & $-$44:20:53.8 & 4955727815260641408 & 2018-10-01 &  40 & 16.99 & 0.7 & 19 & 34 \\
Phoenix 10 & 01:51:02.50 & $-$43:02:41.0 & 4956084950380306816 & 2018-09-30 & 134 & 16.64 & 1.0 & 23 & 41 \\
\enddata
\end{deluxetable*}

We observed most stars in a single epoch. The exceptions were Phoenix-6 and Phoenix-10, which we observed on two different runs. No significant radial velocity differences were present between the two epochs for these stars, so we stacked the observations as if they were conducted in a single night.

\section{Methods}
\label{sec:methods}

We determined the radial velocity of each observation by a cross-correlation of each echelle order against a high-S/N rest-frame spectrum of the very metal-poor star HD122563 \citep{Frebel:2010}. The quoted radial velocity is a mean of individual orders, weighted by the inverse variance of individual velocity estimates \citep{Ji:2020}, with barycentric corrections applied. The radial velocities for Phoenix-6 and Phoenix-10 are a weighted average of individual epochs. 

We continuum-normalised each echelle order using spline functions with saturated lines masked out, and iterative sigma clipping to down-weight other absorption lines. The continuum-normalised spectra from individual orders were then combined, with each resampled pixel weighted by inverse variance, providing a contiguous spectrum from 332\,nm to 941\,nm. The typical S/N per pixel at 400\,nm is 20, and 37 at 650\,nm (Table~\ref{tab:observations}).

We performed a standard chemical abundance analysis assuming 1D plane-parallel \texttt{ATLAS} model atmospheres \citep{Castelli:2004} and local thermal equilibrium (LTE). The 2017 version of the \texttt{MOOG} radiative transfer code with improved treatment of scattering \citep{Sneden:1973,Sobeck:2011}, was used throughout this analysis, wrapped with the \texttt{SMHR} analysis code \citep{Casey:2014}. Equivalent widths of unblended atomic lines were measured from the continuum-normalised spectra, allowing for local continuum fitting around each absorption line. Specifically, we used Gaussian profiles to fit absorption lines, and simultaneously fit the parameters of the Gaussian and a straight line continuum function, while iteratively masking out nearby absorption lines that would otherwise bias our model fit. With each iteration we tested if there were nearby absorption lines that could provide better fit by using a Gaussian profile with the same standard deviation as the absorption line of interest (i.e., same spectral resolution), and if so we excluded the pixels surrounding that nearby absorption feature. Verification tests of this algorithm are shown elsewhere \citep{Casey:2014,Ji:2020}.

The stellar parameters were estimated consistently for all high-resolution spectroscopic observations taken for the $S^5$ survey \citep{Li:2019,Ji:2020}. This allows for a consistent comparison of abundances in different stellar streams. Complete details on the methods to estimate stellar parameters are given in \citet{Ji:2020}, which we summarise here. We estimated effective temperatures using Dartmouth isochrones \citep{Dotter:2008} and de-reddened $g-r$ photometry from the DES \citep{Morganson:2018}. Specifically we use 12 Gyr $\alpha$-element enhanced isochrones with [Fe/H] $= -2.5$, $-1.5$, and $-1.0$. For extinction correction in $g$ and $r$ we adopted the colour excess $E(B-V)$ from \citep{Schlegel:1998} and extinction coefficients from the first DES data release \citep{DESDR1} (see also Equations 1 and 2 from \citet{Ji:2020}). Assuming a distance modulus from \citet{Shipp:2018}, we used the isochrone with the closest predicted de-reddened $g$-band magnitude and converted the de-reddened $g-r$ colour to effective temperature. Uncertainties from photometry, and the choice of isochrone, are propagated to the uncertainty in temperature (about 50-60\,K). We estimated surface gravity $\logg$ from DES photometry \citep{DESDR1}, with bolometric corrections from \citet{Casagrande:2014}, and report a typical uncertainty of 0.16\,dex. Uncertainties in distance modulus, photometry, and temperature are propagated to our uncertainties in surface gravity.

\begin{figure}
	\includegraphics[width=\columnwidth]{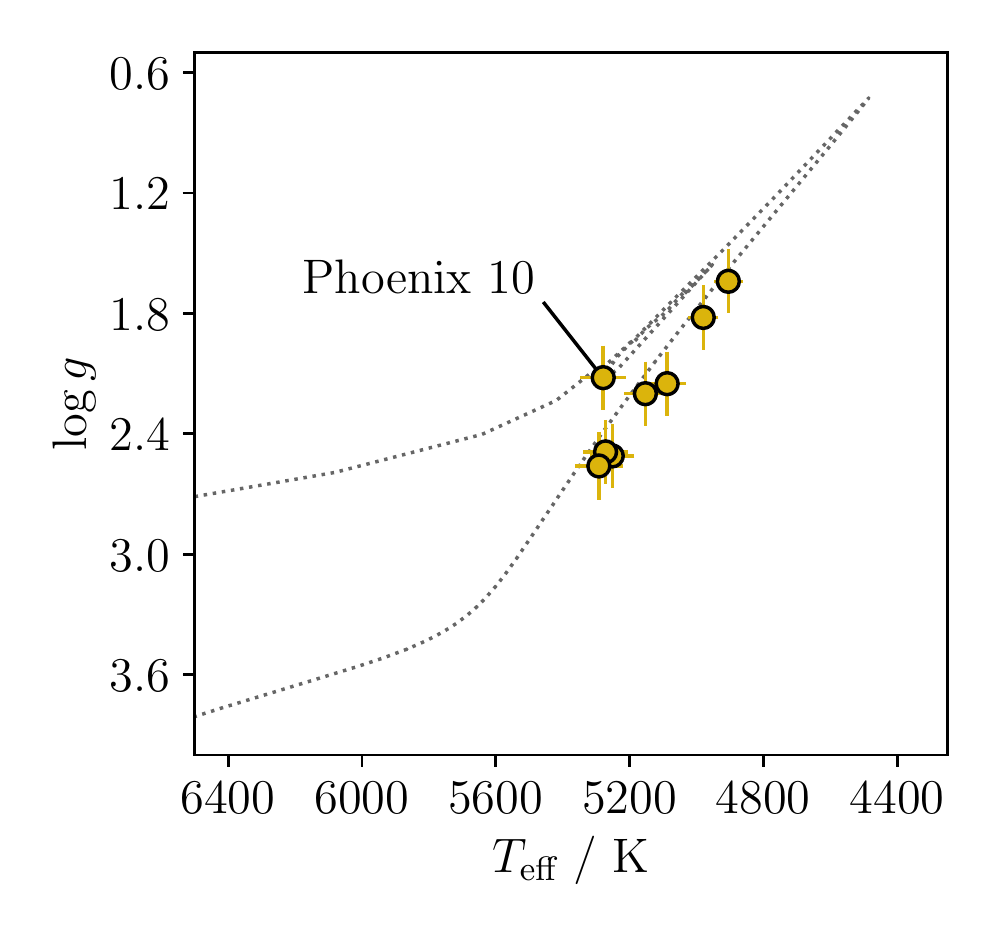}
	\caption{Effective temperature $\teff$ and surface gravity $\logg$ for all Phoenix members observed with Magellan/MIKE, compared to a 12 Gyr metal-poor isochrone \citep{Dotter:2008}. The lithium-rich giant star (Phoenix-10) is marked.\label{fig:hrd}}
\end{figure}

Equivalent widths of unblended \ion{Fe}{2} lines were used to estimate the remaining stellar parameters. With these equivalent width measurements and fixed $\teff$ and $\logg$, we estimated the microturbulence $\vt$ by balancing the \ion{Fe}{2} abundance with respect to the reduced equivalent width, requiring that the mean \ion{Fe}{2} abundance matched the model atmosphere metallicity. We assume $\alpha$-enhanced atmospheres ([$\alpha$/Fe] = +0.4), and a model metallicity uncertainty of 0.2\,dex for all stars. The stellar parameters are given in Table~\ref{tab:stellar-parameters} (see also Figure~\ref{fig:hrd}). The photometry and spectroscopic analysis of all Phoenix stars observed (except one) are consistent with being first ascent red giant branch stars (Figure~\ref{fig:hrd}). The exception is Phoenix-10, which has stellar parameters that are more consistent with being a red clump star.

Detailed chemical abundances are derived from equivalent width measurements for unblended atomic lines, or by spectral synthesis. Uncertainties in chemical abundances include systematic and statistical uncertainties, including correlations between stellar parameters. Upper limits on abundances are estimated by spectral synthesis, given the stellar parameters, an estimated spectral resolution from nearby absorption lines, and the continuum model. We refer the reader to Appendices  B through D of \citet{Ji:2020} for full details of how uncertainties are estimated, including the construction of the line list and verification of our equivalent widths and spectral syntheses. In Table~\ref{tab:chemical-abundances} we list the chemical abundances for each Phoenix star, which are also shown in Figure~\ref{fig:abundances}. \citet{Ji:2020} provides the line-by-line abundances and uncertainties for all stream stars observed as part of $S^5$.

\begin{deluxetable}{lcccc}
\tablecolumns{5}
\tablecaption{\label{tab:stellar-parameters}Stellar parameters for all targets. Note that [M/H] refers to the mean stellar metallicity, derived from \ion{Fe}{2} lines \citep{Ji:2020}.}
\tablehead{Star & $T_\mathrm{eff}$ & $\log{g}$ & $\nu_t$ & [M/H] \\
& (K) & & (km\,s$^{-1}$)}
\startdata
Phoenix 1   & $5088 \pm 57$ & $2.15 \pm 0.16$ & $1.47 \pm 0.19$ & $-2.52$ \\
Phoenix 2   & $5252 \pm 66$ & $2.51 \pm 0.16$ & $1.64 \pm 0.30$ & $-2.67$ \\
Phoenix 3   & $5272 \pm 67$ & $2.49 \pm 0.16$ & $1.49 \pm 0.38$ & $-2.76$ \\
Phoenix 6   & $4905 \pm 43$ & $1.64 \pm 0.16$ & $2.11 \pm 0.59$ & $-2.68$ \\
Phoenix 7   & $4980 \pm 45$ & $1.82 \pm 0.16$ & $1.58 \pm 0.18$ & $-2.62$ \\
Phoenix 8   & $5292 \pm 71$ & $2.56 \pm 0.17$ & $1.53 \pm 0.07$ & $-2.79$ \\
Phoenix 9   & $5153 \pm 64$ & $2.20 \pm 0.16$ & $1.55 \pm 0.27$ & $-2.70$ \\
Phoenix 10  & $5279 \pm 68$ & $2.12 \pm 0.16$ & $1.80 \pm 0.33$ & $-2.93$ \\
\enddata
\end{deluxetable}

\begin{deluxetable*}{llccrrrrrrrrrr}
\tablecolumns{14}
\tabletypesize{\footnotesize}
\tablecaption{\label{tab:chemical-abundances}Summary of stellar abundances, including  differences due to a 1$\sigma$ change in stellar parameters.}
\tablehead{Star & Species & $N$ & ul & $\log \epsilon$ & [X/H] & $\sigma_{\text{[X/H]}}$ & [X/Fe] & $\sigma_{\text{[X/Fe]}}$ & $\Delta_T$ & $\Delta_g$ & $\Delta_v$ & $\Delta_M$ & $s_X$}
\startdata
Phoenix 1     & C-H   &   2 &     &$+6.17$&$-2.26$&  0.18 &$+0.34$&  0.18 &  0.13 & -0.28 &  0.01 &  0.04 &  0.03 \\
Phoenix 1     & C-N   &   1 & $<$ &$+5.78$&$-2.05$&\nodata&$+0.54$&\nodata&\nodata&\nodata&\nodata&\nodata&\nodata\\
Phoenix 1     & O I   &   1 & $<$ &$+8.04$&$-0.65$&\nodata&$+1.95$&\nodata&\nodata&\nodata&\nodata&\nodata&\nodata\\
Phoenix 1     & Na I  &   2 &     &$+3.94$&$-2.30$&  0.13 &$+0.29$&  0.12 &  0.05 & -0.22 & -0.09 & -0.02 &  0.06 \\
Phoenix 1     & Mg I  &   7 &     &$+5.38$&$-2.23$&  0.05 &$+0.37$&  0.06 &  0.04 & -0.07 & -0.02 &  0.01 &  0.06 \\
Phoenix 1     & Al I  &   2 &     &$+3.42$&$-3.03$&  0.30 &$-0.43$&  0.30 &  0.08 & -0.26 & -0.09 &  0.01 &  0.30 \\
Phoenix 1     & Si I  &   2 &     &$+5.46$&$-2.04$&  0.10 &$+0.55$&  0.10 &  0.07 & -0.12 & -0.04 &  0.01 &  0.00 \\
Phoenix 1     & K I   &   2 &     &$+3.01$&$-2.02$&  0.09 &$+0.58$&  0.09 &  0.04 & -0.10 & -0.03 & -0.00 &  0.05 \\
Phoenix 1     & Ca I  &  13 &     &$+4.03$&$-2.31$&  0.04 &$+0.29$&  0.05 &  0.04 & -0.08 & -0.03 &  0.00 &  0.09 \\
Phoenix 1     & Sc II &   6 &     &$+0.56$&$-2.59$&  0.15 &$-0.07$&  0.17 &  0.04 &  0.04 &  0.01 &  0.02 &  0.16 \\
Phoenix 1     & Ti I  &  11 &     &$+2.62$&$-2.33$&  0.07 &$+0.27$&  0.07 &  0.07 & -0.12 & -0.03 &  0.01 &  0.18 \\
Phoenix 1     & Ti II &  28 &     &$+2.75$&$-2.19$&  0.13 &$+0.33$&  0.11 &  0.04 &  0.11 &  0.05 &  0.03 &  0.17 \\
Phoenix 1     & V I   &   1 & $<$ &$+1.80$&$-2.13$&\nodata&$+0.47$&\nodata&\nodata&\nodata&\nodata&\nodata&\nodata\\
Phoenix 1     & V II  &   1 & $<$ &$+1.98$&$-1.95$&\nodata&$+0.57$&\nodata&\nodata&\nodata&\nodata&\nodata&\nodata\\
Phoenix 1     & Cr I  &   2 &     &$+2.61$&$-3.03$&  0.08 &$-0.44$&  0.08 &  0.07 & -0.11 & -0.02 &  0.01 &  0.00 \\
Phoenix 1     & Fe I  &  81 &     &$+4.90$&$-2.60$&  0.04 &$+0.00$&  0.00 &  0.06 & -0.10 & -0.04 &  0.01 &  0.23 \\
Phoenix 1     & Fe II &  12 &     &$+4.98$&$-2.52$&  0.16 &$+0.00$&  0.00 &  0.02 &  0.13 &  0.00 &  0.01 &  0.08 \\
Phoenix 1     & Co I  &   3 &     &$+2.61$&$-2.38$&  0.11 &$+0.21$&  0.12 &  0.06 & -0.07 & -0.01 &  0.00 &  0.10 \\
Phoenix 1     & Ni I  &   5 &     &$+3.66$&$-2.56$&  0.08 &$+0.04$&  0.09 &  0.06 & -0.10 & -0.05 &  0.00 &  0.10 \\
Phoenix 1     & Cu I  &   1 & $<$ &$+2.42$&$-1.77$&\nodata&$+0.82$&\nodata&\nodata&\nodata&\nodata&\nodata&\nodata\\
Phoenix 1     & Zn I  &   2 &     &$+2.31$&$-2.25$&  0.11 &$+0.35$&  0.12 &  0.03 &  0.03 & -0.01 &  0.01 &  0.00 \\
Phoenix 1     & Sr II &   2 &     &$+0.51$&$-2.36$&  0.24 &$+0.16$&  0.22 &  0.00 &  0.03 & -0.08 & -0.04 &  0.18 \\
Phoenix 1     & Ba II &   3 &     &$-1.41$&$-3.58$&  0.16 &$-1.06$&  0.14 &  0.04 &  0.07 & -0.01 &  0.01 &  0.10 \\
Phoenix 1     & La II &   1 & $<$ &$-0.58$&$-1.68$&\nodata&$+0.84$&\nodata&\nodata&\nodata&\nodata&\nodata&\nodata\\
Phoenix 1     & Eu II &   2 & $<$ &$-1.43$&$-1.95$&\nodata&$+0.57$&\nodata&\nodata&\nodata&\nodata&\nodata&\nodata\\
\enddata
\tablecomments{One star from this table is shown to demonstrate its form and content. The full version is available online.}
\end{deluxetable*}

Using measurements and upper limits of chemical abundances for all observed Phoenix stars, we modelled the intrinsic scatter in abundances for each chemical element. We assumed that the Phoenix stream has an unknown abundance mean $\mu$, some intrinsic scatter $\sigma$, and that the observed abundance of each star is drawn from a normal distribution $\mathcal{N}(\mu,\sigma)$. We  assumed that the true abundance of each star is not known, and that each observation is a noisy realisation of the unknown true abundance. The estimated uncertainty in each abundance measurement includes random and systematic effects. If a star has an upper limit then we assume that the true abundance for that star is not known, and is uniformly drawn between $[\textrm{X/H}] \sim \mathcal{U}(-4, \hat{u})$, where $\hat{u}$ is the reported limit value. We adopt an improper uniform prior on $\mu$ and a uniform prior on $\log{\sigma} \sim \mathcal{U}\left(-5, 0\right)$. We used the \texttt{Stan} \citep{Carpenter:2017,Salvatier:2016} implementation of a Hamiltonian Monte Carlo sampler to draw posterior samples of $\sigma$ and other nuisance parameters (e.g., the unknown true abundance when only limits are available) for each chemical element. 

\section{Results} 
\label{sec:results}

The stellar parameters we report confirm the results by \citet{Wan:2020} from low-resolution spectra of 11 red giant branch stars with $\mathrm{S/N}~>~10\,\mathrm{pixel}^{-1}$. Our results are based on high-dispersion spectra of 8 of those stars. \citet{Wan:2020} use equivalent width measurements of the infrared \ion{Ca}{2} triplet lines, and the absolute magnitude in the $V$ band, to estimate the stellar metallicity \citep{Carrera:2013}. The distance to the stream is used to calculate the absolute magnitude, implying that the calculation is only valid for genuine stream members. Our analysis does not rely on this assumption, and the results between both studies are in excellent agreement.

\citet{Wan:2020} estimate the mean metallicity of the stream to be $[\mathrm{Fe/H}] = -2.70 \pm 0.06$, and we find the mean metallicity to be $\meanFeIIH$. The agreement for both the mean metallicity and its uncertainty is excellent given the different approaches employed to measure metallicities.
Similarly, \citet{Wan:2020} estimate the intrinsic spread in metallicity in the stream to be $\sigma(\mathrm{[Fe/H]}) = 0.07^{+0.07}_{-0.05}$, and we find $\spreadFeIIH$. The intrinsic metallicity spread remains consistent with zero, and the relatively high-tailed uncertainty in our posterior distribution is related to our prior on $\log\sigma \sim \mathcal{U}\left(-5, 0\right)$. 

\begin{figure*}
	\includegraphics[width=\textwidth]{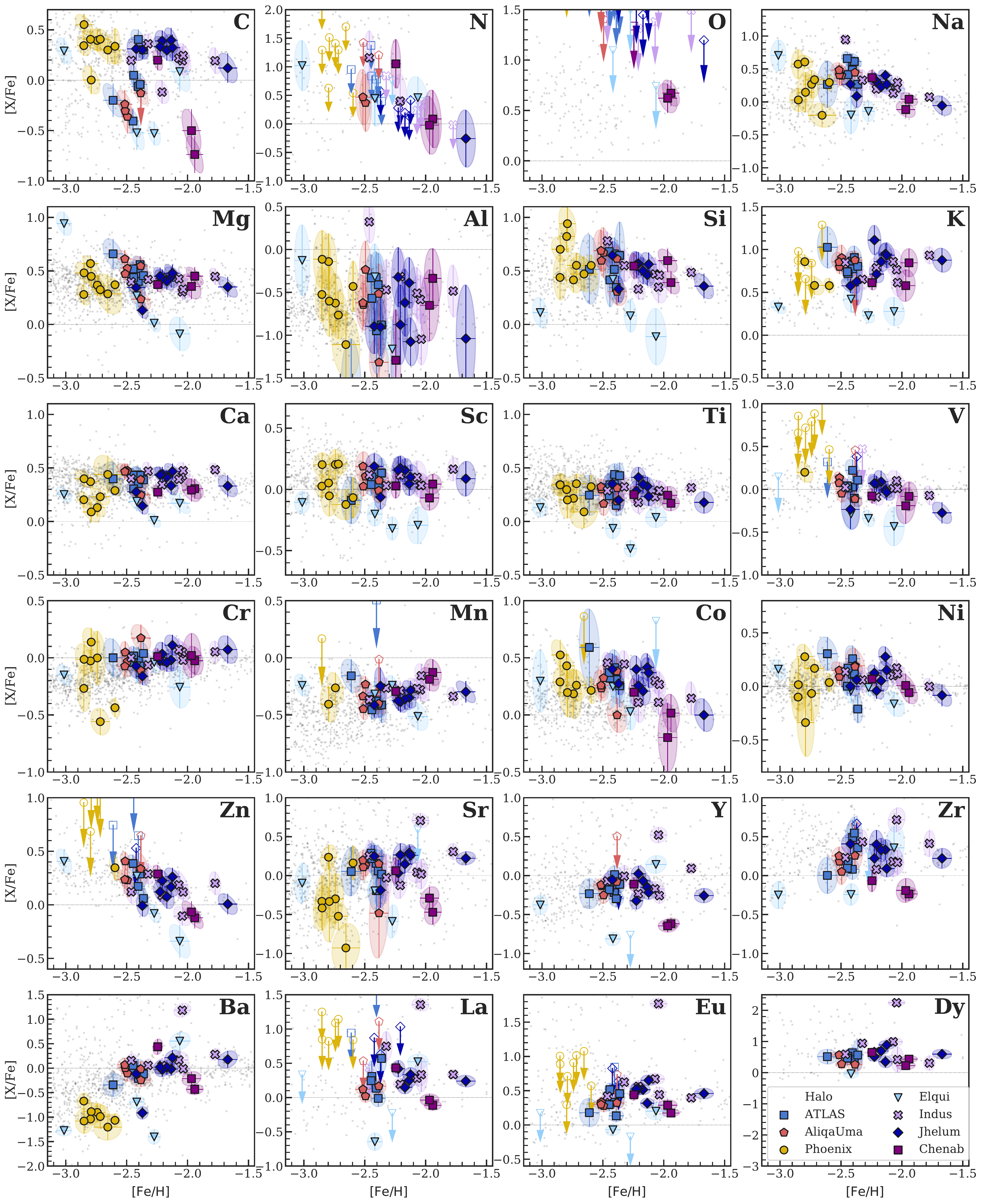}
	\caption{Element abundance ratios for all Phoenix stream stars compared to a literature compilation of Milky Way halo stars \citep[light grey;][]{Abohalima:2018, Fulbright:2000, Barklem:2005, Aoki:2009, Cohen:2013, Roederer:2014c}, and to all other streams with high-resolution spectra acquired through the $S^5$ survey \citep{Ji:2020}. The errors include random and systematic effects. Arrows indicate upper limits.  Figure originally from \citet{Ji:2020}.\label{fig:abundances}}
\end{figure*}

The chemical abundances for the Phoenix stream are shown in Figure~\ref{fig:abundances}, where we compare to a sample of Milky Way stars, and other stream stars with high-resolution spectra from the $S^5$ program \citep{Li:2019, Ji:2020}. The Phoenix stream abundance ratios are comparable to the Milky Way halo in most element abundances. Note the small spread in most abundances relative to other streams, particularly Elqui, which is of a comparable overall mean metallicity. 

\begin{figure}
	\includegraphics[width=\columnwidth]{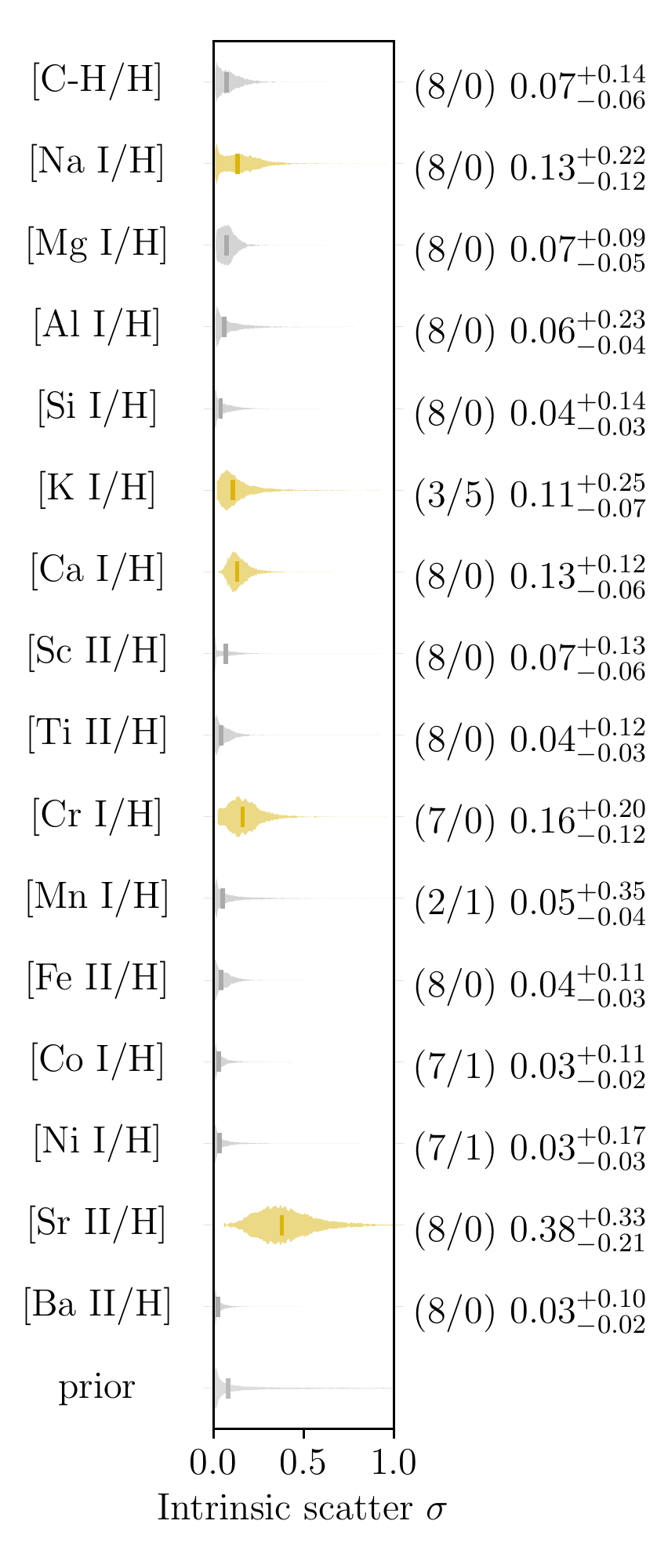}
	\caption{Intrinsic scatter in elemental abundances for the Phoenix stream. Each violin plot visualises the posterior distribution of intrinsic scatter for that chemical element. The vertical line indicates the median value. 
	On the right hand side we show the (number of measurements/number of upper limits) next to the maximum a posteriori (MAP) values, as well as the 5th and 95th percentiles. For elements with only few measurements (e.g., Mn) we recover the prior (also shown) of $\log\sigma \sim \mathcal{U}\left(-5, 0\right)$. )  Elements shaded yellow are those with a MAP intrinsic scatter value more than 0.10\,dex. \label{fig:intrinsic-scatter}}
\end{figure}

We show the posterior distribution for intrinsic scatter in each chemical element in Figure~\ref{fig:intrinsic-scatter}. The level of intrinsic scatter varies per element, with some elements being consistent with no scatter (e.g., Fe II, Ba II), while others exhibit large and significant scatter (e.g., Sr). Before discussing the element with the most significant scatter (Sr), we provide some context here to clarify how to interpret Figure~\ref{fig:intrinsic-scatter}. The posterior in $\spreadMnIH$ suggests a long high-end tail, as judged by the high $+0.35$ positive uncertainty, but this is informed by just two measurements and one upper limit. It is clear at the high end we are only recovering the prior $\log{\sigma} \sim \mathcal{U}\left(-5, 0\right)$, and we cannot confidently rule out large intrinsic scatter given the data. This argument also applies to K (positive uncertainty of $+0.25$\,dex), where only 3 measurements and 5 upper limits exist.

Al~I provides another example to clarify interpretation. The peak-to-peak abundance range in [Al~I/H] is large ($\approx$0.8\,dex; or 1 dex in [Al~I/Fe]) but the reported intrinsic scatter in [Al\,I/H] is low: $\spreadAlIH$. This is because the estimated uncertainties on individual Al~I measurements is larger than other elements, making the distribution of values consistent with very little intrinsic scatter, and in line with the quoted uncertainties. For these reasons, the peak-to-peak range of values and the estimated intrinsic spread are informative in slightly different ways. A handful of other elements show mild evidence of non-zero intrinsic scatter: Na ($1\sigma$), K ($1.5\sigma$), Ca ($2.1\sigma$), and Cr ($1.3\sigma$). Even among the most significant of these, Ca, the actual level of intrinsic scatter is still low, $\spreadCaIH$, particularly for globular clusters.

\begin{figure*}
	\includegraphics[width=\textwidth]{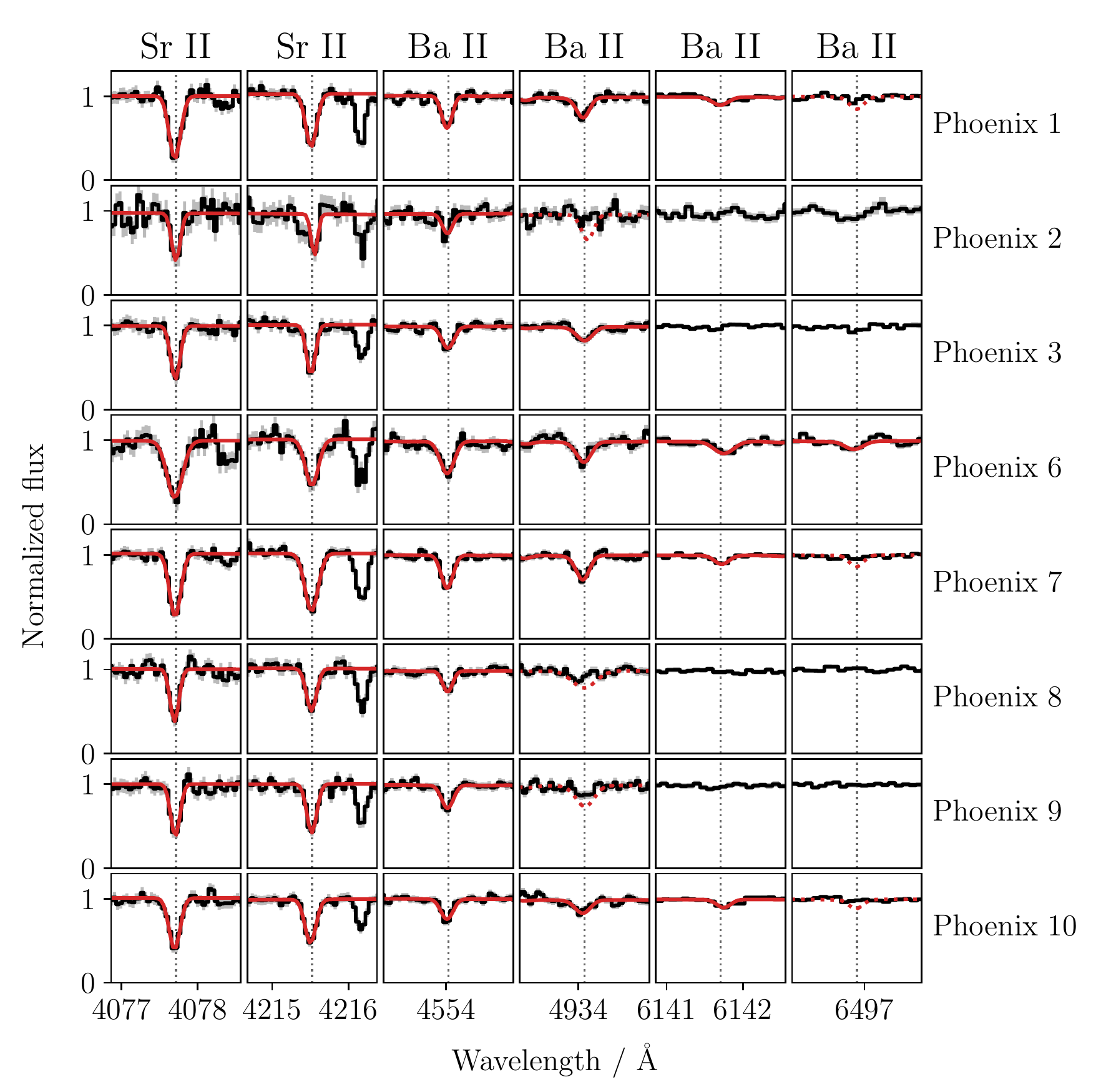}
	\caption{Continuum-normalized spectra surrounding Sr~II and Ba~II transitions in Phoenix stars (black). Spectral fits shown in solid red indicate measurements, and dashed red lines indicate the model was used as an upper limit. Upper limits on individual line abundances are only used if no measurements were available, which is not the case here, but for completeness we have included these upper limits that we used to ensure abundances were consistent given the line list and stellar parameters. Stars Phoenix 1 and 7 have the highest [Sr~II/Fe] ratios while Phoenix 2 has the lowest (see Table~\ref{tab:chemical-abundances}).\label{fig:sr-ba-spectrum-fits}}
\end{figure*}

The element with the largest estimated intrinsic scatter is Sr, with $\spreadSrIIH$. The peak-to-peak range of [Sr\,II/H] values is a little more than $\approx$1\,dex, but the uncertainties on individual measurements are many times smaller than this range, implying some level of intrinsic scatter in the stream. Indeed, some stars are relatively Sr-rich, with $[\mathrm{Sr/Fe}] \sim +0.2$ at $[\mathrm{Fe/H}] \sim -2.6$. Contrast this with Ba, another neutron-capture element that also has strong absorption lines in metal-poor stars. Naively we might predict the spread in Ba and Sr to be comparable given the individual abundance uncertainties. Instead we find that the mean [Ba\,II/H] is very low ($[\mathrm{Ba/Fe}] \sim -1$) and the stream has negligible intrinsic scatter for this element. In comparison, the abundance ratios of [Sr\,II/H] show a reasonably large spread ($\gtrsim$1\,dex) and are inconsistent with zero scatter at the $\approx2\sigma$ level. We show the spectral fits of Ba and Sr transitions in Figure~\ref{fig:sr-ba-spectrum-fits}.

The [Sr\,II/H] abundance ratios we measure show no correlations with stellar parameters that would be consistent with departures from local thermodynamic equilibrium (LTE) or other missing systematic effects. Indeed, \citet{Hansen:2013} estimate the non-LTE departure coefficients for metal-poor ($[\mathrm{Fe/H}] \approx -2.8$) giants to be $<\pm0.05$\,dex for the 407.77\,nm absorption line used in this work, and at most 0.2\,dex across all lines \citep{Andrievsky:2011}.\footnote{The non-LTE correction for the 421.55\,nm \ion{Sr}{2} line, a line which we used in this work, is not computed by \citet{Hansen:2013} due to blended Fe lines that affect the metal-rich stars in that work. In this work we find no significant difference in abundance between the 407.77\,nm and 421.55\,nm lines, of which the former has no significant correction due to non-LTE effects.}  We conclude that the intrinsic abundance spread we infer in Sr appears \emph{bona fide}.

Only upper limits are available for all other neutron-capture elements. We find limits on [Eu/Fe] ranging between 0 and $+1$, and $\lesssim 0.82$ is our strongest limit on [La/Fe]. As the Phoenix stream stars are not clearly enhanced in r-process elements (e.g., Eu), we do not report limits on other neutron-capture elements. In particular, even among the stars with highest [Sr/Fe] abundance ratios, we do not have useful limits on other light s-process elements like Y or Zr: only one (uninformative) upper limit is available.

\begin{figure}
	\includegraphics[width=\columnwidth]{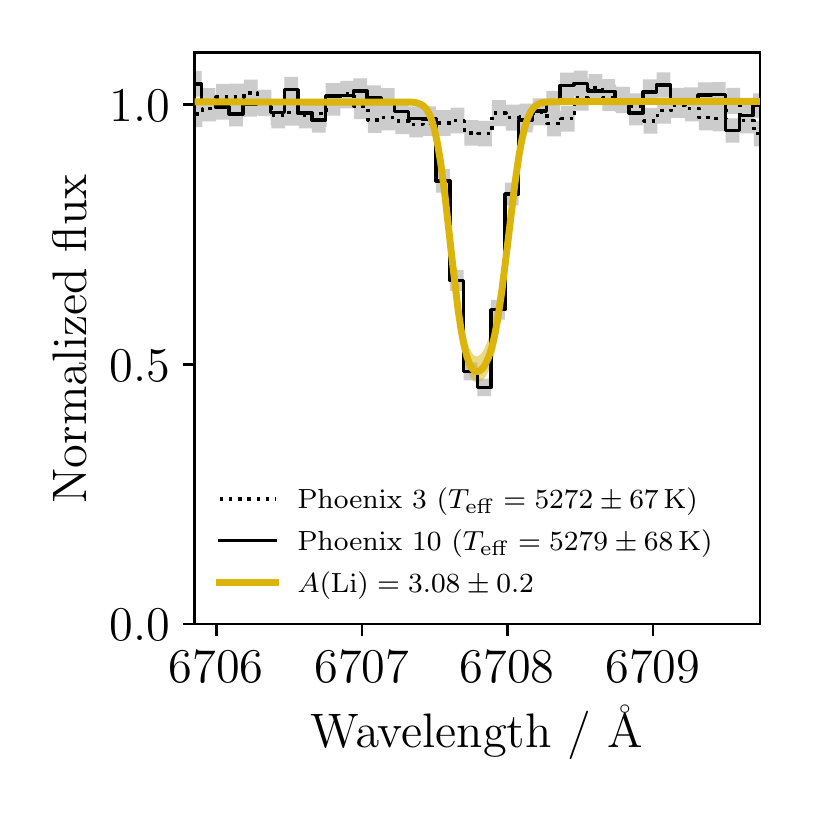}
	\caption{High-resolution spectrum of Phoenix-10 (the lithium-rich giant, with yellow showing the best-fitting model; $A(\textrm{Li}) = 3.1 \pm 0.2\,\textrm{dex}$) centered at the lithium doublet at 6707\,\AA\ compared to Phoenix-3, a lithium-normal star of comparable temperature.\label{fig:li-rich}}
\end{figure}

The remaining chemical abundance worthy of mention is the enhancement of lithium in Phoenix-10. With $A(\mathrm{Li}) = 3.1 \pm 0.2$, Phoenix-10 easily meets the classification of a so-called lithium-rich giant star.\footnote{$A(\mathrm{Li}) > 1.5$ is a common definition.} The stellar parameters (and relative lifetimes of evolutionary phases) make Phoenix-10 more likely to be a core-helium burning star than a red giant branch star, but we have no asteroseismic data to firmly distinguish these scenarios. Regardless of its evolutionary state, the observed lithium abundance exceeds what is expected for evolved stars. Figure~\ref{fig:li-rich} shows the lithium doublet at 6707\,\AA\ for Phoenix-10 with a best-fitting model, and the spectrum of Phoenix-3, a giant star of nearly identical temperature and no detectable lithium absorption. Phoenix-10 shows no other peculiar chemical abundances compared to other Phoenix stream stars, typical for lithium-rich giants \citep{Casey:2016}.

\begin{figure}
	\includegraphics[width=\columnwidth]{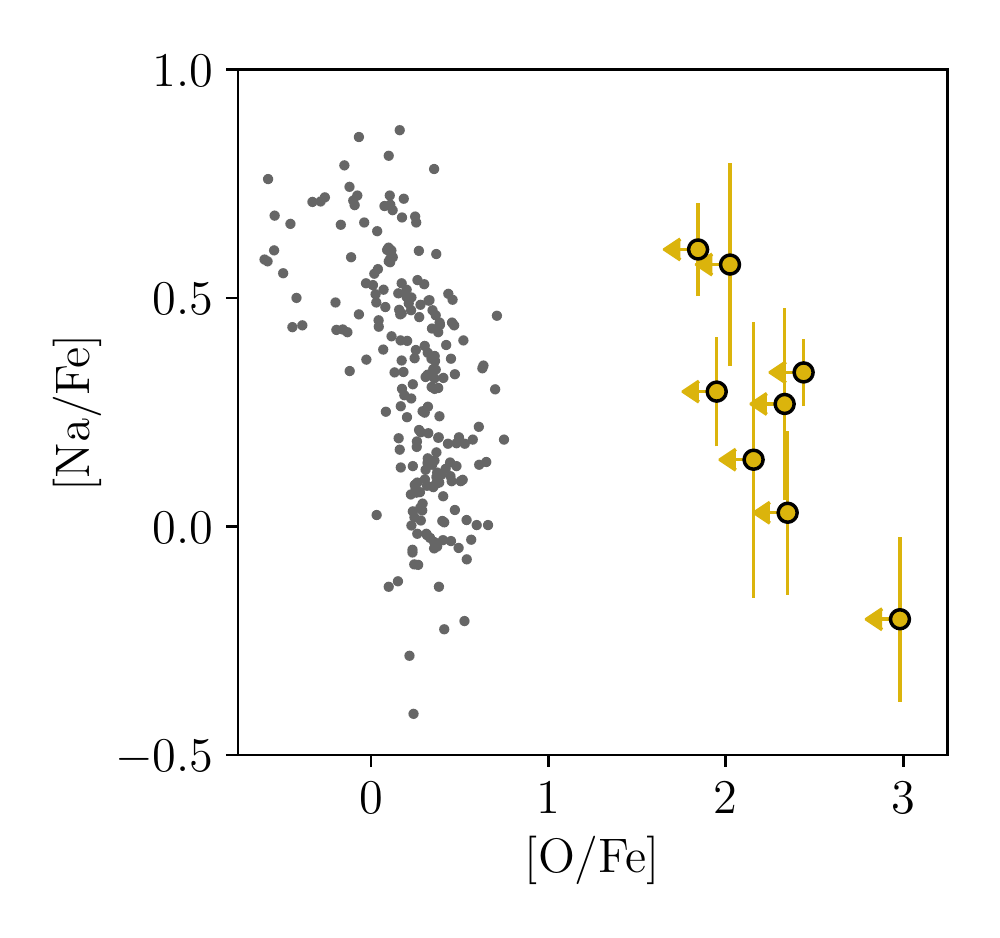}
	\caption{[O/Fe] and [Na/Fe] abundances for globular clusters \citep[dark grey;][]{Carretta:2009} compared to the Phoenix stream (yellow). \label{fig:na-o}}
\end{figure}

\begin{figure}
	\includegraphics[width=\columnwidth]{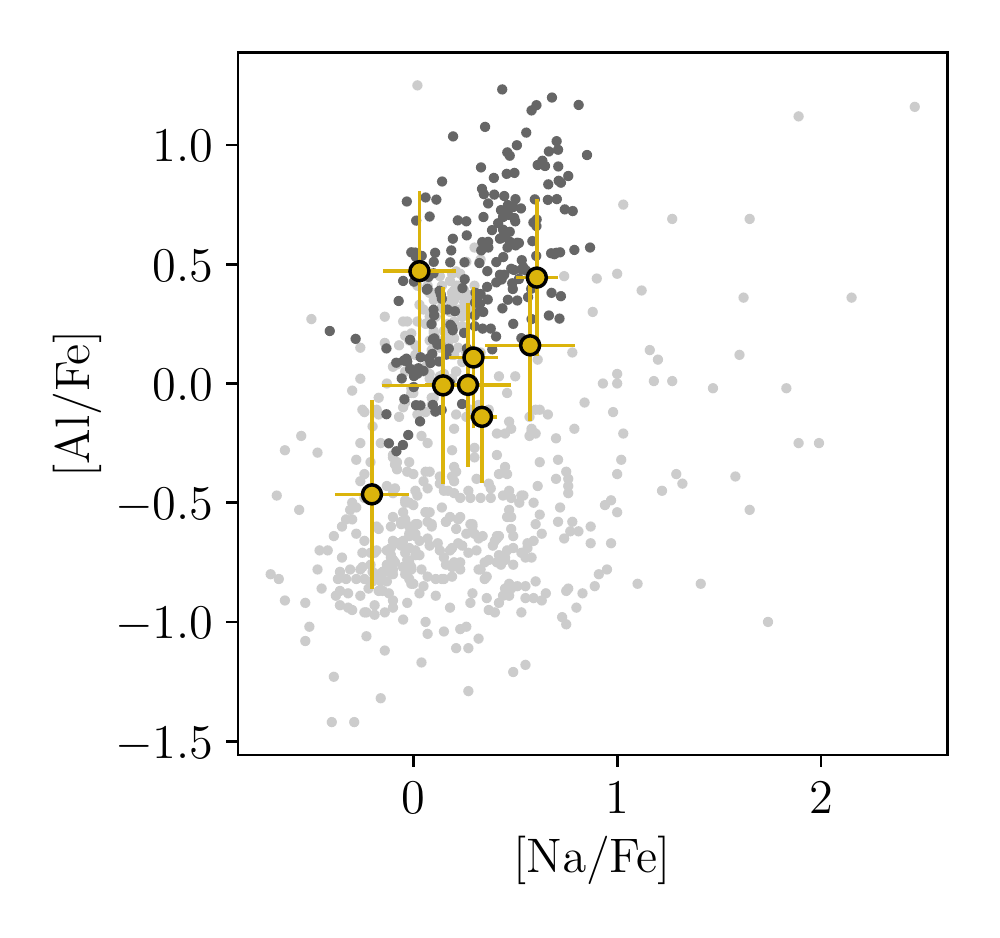}
	\caption{[Na/Fe] and [Al/Fe] abundances for the Phoenix stream stars (yellow) compared to literature globular cluster measurements \citep[dark grey;][]{Carretta:2009} and Milky Way halo stars \citep[light grey;][]{Abohalima:2018, Fulbright:2000, Barklem:2005, Aoki:2009, Cohen:2013, Roederer:2014c}. Note that the [Al/Fe] abundance ratios shown here for the Phoenix stream have had non-LTE corrections applied on a per-line basis \citep{Nordlander:2017} to be consistent with the literature values (see Section~\ref{sec:discussion}).\label{fig:na-al}}
\end{figure}

\section{Discussion} 
\label{sec:discussion}

The low intrinsic scatter in metallicity we find in the Phoenix stream is consistent with the low intrinsic velocity spread, and the narrow stream width. These results imply that the former Phoenix star cluster was a low-mass ($M \sim 3 \times 10^4 M_\odot$) metal-poor globular cluster, in agreement with the previous low-resolution analysis \citep{Wan:2020}. The low intrinsic scatter in metallicity strongly disfavours an ultra faint dwarf galaxy classification, where  spreads of 0.2-0.3\,dex are common.

The Phoenix stream has a comparable orbital energy and azimuthal action to the Palomar~5 stream, the metal-poor globular cluster NGC~5053, and the Helmi stream \citep{Wan:2020}.  Despite these orbital similarities, there appears to be no definitive connection in chemical abundances. The mean metallicity of Palomar~5 is $[\mathrm{Fe/H}] = -1.48\,\pm\,0.10$ \citep{Kuzma:2015}, distinct from Phoenix at $[\mathrm{Fe/H}] \approx -2.7$. NGC~5053 is closer in overall metallicity \citep[$-2.30$;][]{Carretta:2009}, but still significantly different by 0.4\,dex. The Helmi stream also appears unrelated, as it demonstrates a very wide spread in metallicities ($-2.3$ to $-1.0$, \citealt{Koppelman:2019}), that is indicative of a dwarf galaxy origin. We note that the Phoenix stream is also spatially aligned with the Hermus stream, and while there is limited abundance information for the Hermus stream, the orbits of the two streams appear inconsistent with each other \citep{Martin:2019}.


We find the Phoenix stream chemical abundances are consistent with a globular cluster classification. All globular clusters show an anti-correlation between Na and O abundance ratios that is dominated by large spreads in [Na/Fe] with a depletion of [O/Fe] at the highest [Na/Fe] levels \citep[e.g.,][]{Carretta:2019}. We find [Na/Fe]\footnote{In this work [X/Fe] and [X/H] are interchangeable given the low intrinsic scatter in [Fe/H].} in the Phoenix stream ranges from $-0.20$ to $+0.61$, nearly spanning the full range observed in globular clusters (Figure~\ref{fig:na-o}). Unfortunately, we have only upper limits on [O/Fe] from the 630\,nm absorption line, which are largely uninformative: they prevent us from detecting the presence (or absence) of any Na-O abundance correlation, despite the spread in [Na/Fe]. 

The Mg-Al abundance pattern in globular clusters can be similarly described as a large abundance spread in [Al/Fe] and a small spread in [Mg/Fe], with decreasing [Mg/Fe] ratios at the highest [Al/Fe] ratios. While the Na-O pattern appears to be present in every globular cluster, the Mg-Al abundance pattern is either less apparent or non-existent in lower-mass or high-metallicity globular clusters \citep{Pancino:2018, Nataf:2019}. For this reason, we may not expect to see any Mg-Al relationship in a low mass system like Phoenix, even if it were a bound globular cluster. However, globular clusters with strong Al enrichment also demonstrate a correlation between [Na/Fe] and [Al/Fe] abundances. 

We find a comparable correlation in the Phoenix stream abundances. In Figure~\ref{fig:na-al} we show these abundance ratios compared to Milky Way halo stars and globular cluster stars reported in the literature. This comparison warrants some discussion. We estimate [Al~I/H] abundance ratios from the 3944\,\AA\ and the 3961\,\AA\ Al~I absorption lines because these features are strong even in very metal-poor stars. However, most literature studies of globular clusters derive Al abundances from the 6696--6698\,\AA\ doublet, including those literature abundances shown in Figure~\ref{fig:na-al}. The 6696--6698\,\AA\ doublet would be preferable to use as is not strongly affected by non-LTE effects, but it is not visible in our spectra due to these lines being generally weaker, and because of the low S/N ratio of our spectra. Instead we are forced to use the 3944\,\AA\ and 3961\,\AA\ absorption lines, but these lines \emph{are} strongly affected by non-LTE in very metal-poor stars \citep{Nordlander:2017}. For this reason, it is prudent to apply a non-LTE correction to our [Al/Fe] abundances. By accounting for the corrections on each line individually, in each star the total correction to the abundance varies between $+0.54$ and $+0.69$\,dex. The corrected [Al/Fe] abundances in the Phoenix stream are consistent with literature studies of globular clusters, and the [Al/Fe] abundance ratios are correlated with [Na/Fe] abundances\footnote{These abundance ratios remain correlated even without the non-LTE corrections applied.}, as seen in other globular clusters.

The neutron-capture elements we measure in the Phoenix stream set it distinctly apart from typical Milky Way halo stars. In Figure~\ref{fig:abundances} we can see that the Phoenix stream has a very small range of [Fe/H] and [Ba/Fe] abundance ratios compared to other streams. ATLAS and Aliqa Uma are also concentrated in this plane and are only slightly more metal-rich than Phoenix, but their abundance ratios are more typical of what is observed in the Milky Way halo. Indeed, Aliqa Uma and ATLAS have a mean [Sr,Ba/Fe] $\sim 0$ and a small spread in both [Sr/Fe] and [Ba/Fe] abundance ratios. Comparing this with the Phoenix stream we find a low [Ba/Fe] $\sim -1$, coupled with a very small spread in metallicity, and large range in [Sr/Fe] abundance ratios. The [Ba/Fe] ratio for the Phoenix stars is a bit low compared to the field halo at the same metallicity, and low compared to metal-poor globular clusters, where [Ba/Fe] is typically $\approx -0.1$. This highlights the differences in Phoenix compared to other stellar streams, in a way that has not been easily possible before due to systematic differences in how chemical abundances are estimated between studies. 

\begin{figure}
	\includegraphics[width=\columnwidth]{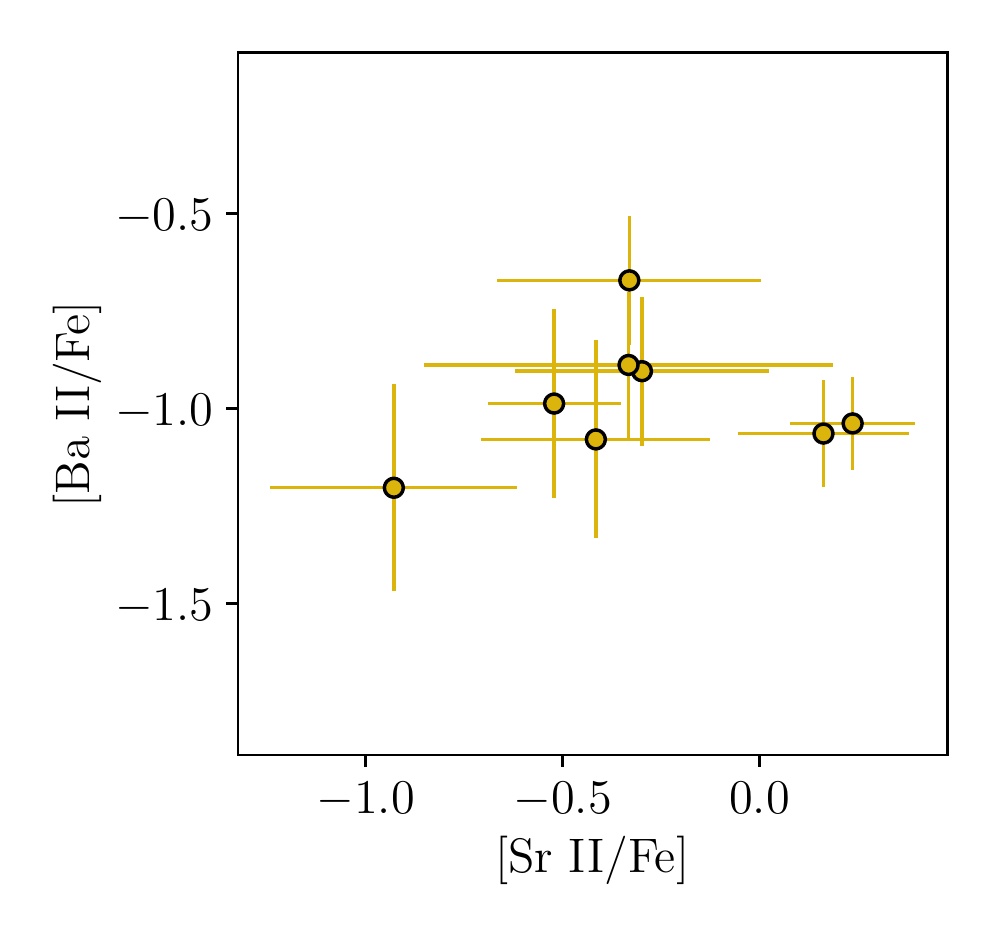}
	\caption{[Ba/Fe] and [Sr/Fe] abundance ratios for Phoenix stream stars, shown with equal-axis limits. \label{fig:ba-sr}}
\end{figure}

We find that strontium shows the largest intrinsic scatter among all elements: $\spreadSrIIH$\,dex (see also Figure~\ref{fig:ba-sr}).  Our model for intrinsic scatter includes estimated random and dominant systematic uncertainties in abundance measurements, making it unlikely that the Sr abundance spread is a result of underestimated uncertainties in individual measurements. There are globular clusters known to show ranges in chemical elements like Sr, but these abundances are generally associated with ranges in overall metallicity (e.g., Fe-rich stars tend to be Sr-rich), and other neutron-capture elements like Ba \citep[e.g.,][]{Marino:2011,Sobeck:2011,Yong:2009,Yong:2014}. To our knowledge, there is no globular cluster that shows a large intrinsic spread in Sr, without an accompanying spread in overall metallicity or other neutron-capture elements. 

Despite the rarity of metal-poor stars with high [Sr/Fe] and low [Ba/Fe] abundance ratios, the abundance signature is clearly not restricted to the Phoenix stream. ROA 276 is a metal-poor star ($[\mathrm{Fe/H}] \approx -1.3$) in the globular cluster $\omega$-Centauri that has an unusually high [Sr/Ba] abundance ratio \citep{Stanford:2010}, with high [X/Fe] ratios for all elements from Cu to Mo, and normal ratios for Ba to Pb \citep{Yong:2017}. The best explanation for those abundance ratios was found to be enrichment from a metal-poor ($[\mathrm{Fe/H}] = -1.8$) fast rotating massive-star model of 20\,$M_\odot$, with no other nucleosynthetic source able to match the observed neutron-capture abundances. Similarly, \citet{Jacobson:2015} reported on SMSS~J022423-573705, an extremely metal-poor ($[\mathrm{Fe/H}] = -3.97$) star with $[\mathrm{Sr/Fe}] \sim 1$ and a very constraining limit of $[\mathrm{Ba/Fe}] < 0.91$. 

\begin{figure*}
	\includegraphics[width=\textwidth]{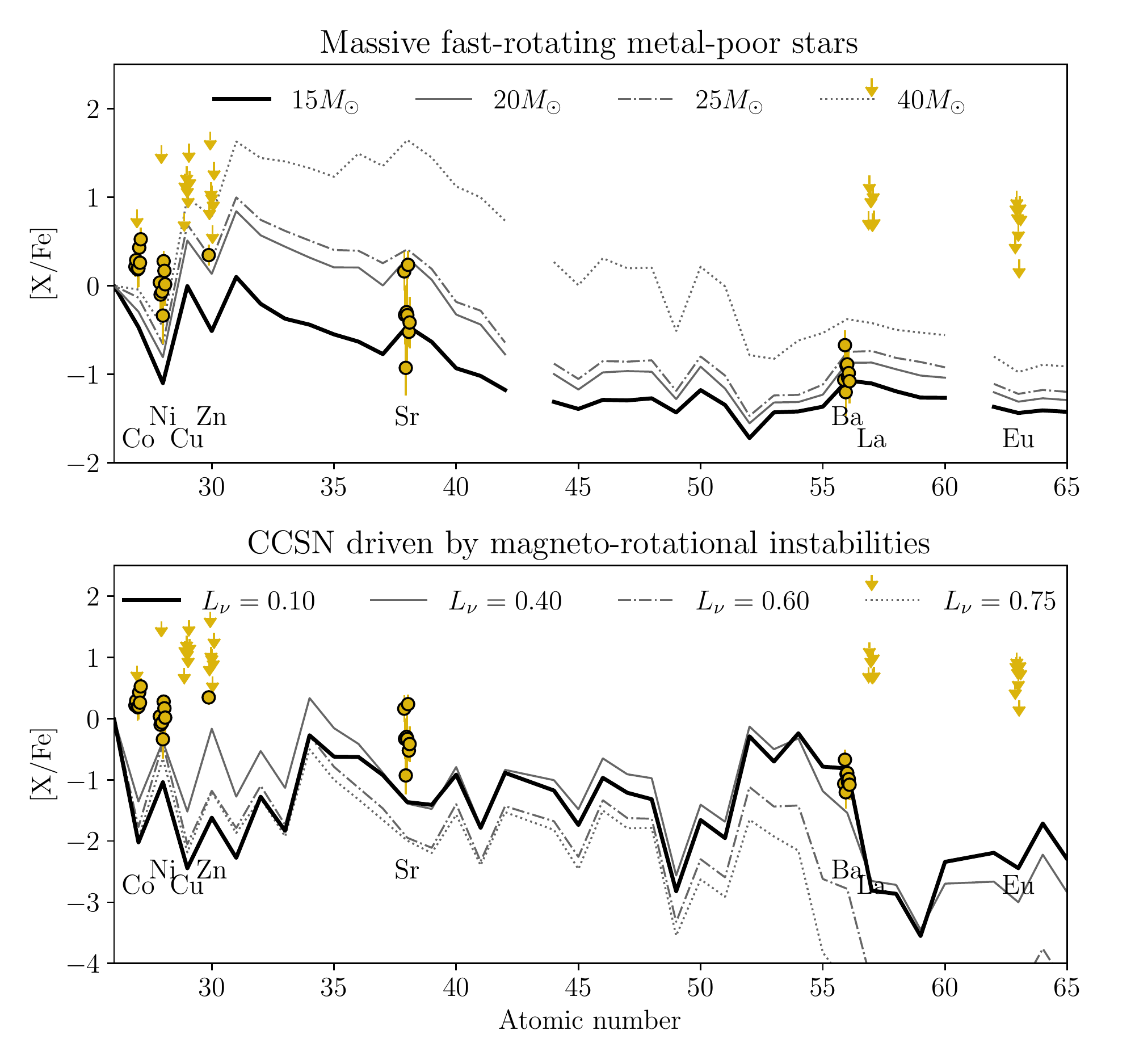}
	\caption{Predicted yields for two scenarios: massive fast-rotating metal-poor stars where neutron-capture elements are produced by rotationally-driven s-process production \citep[top;][]{frischknecht:2012,frischknecht:2016}, and core-collapse supernovae that are driven by a magneto-rotational instability \citep{Nishimura:2017}. The massive fast-rotating metal-poor star predictions are for $[\mathrm{Fe/H}] = -3.8$ with a rotation speed $\vinicrit = 0.4$. The lower panel shows yield predictions for models with varying magnetic-driving strengths: the $\hat{L}_\nu = 0.10$ has the strongest influence of magnetic fields, the $\hat{L}_\nu = 1$ model (not shown) is dominated by neutrino heating, and $\hat{L}_\nu \approx 0.40$ to $0.75$ are intermediate models between the two extrema. The observed abundance ratios are normalized to [\ion{Fe}{1}/H].\label{fig:yield-predictions}}
\end{figure*}

At low metallicities strontium is produced by both the slow-neutron capture process (s-process) and the rapid-neutron capture process (r-process). The net products of the s-process do not appear to be high at low metallicities for three reasons.
First, at low metallicity the primary neutron source for the s-process is through the $^{22}\mathrm{Ne}(\alpha,\mathrm{n})^{25}\mathrm{Mg}$ reaction, and the rate of this reaction decreases at low metallicity. Secondly, the total number of neutron seeds (Fe) also decreases at low metallicity. Finally, the number of neutron poisons (e.g., $^{16}$O) remains constant with metallicity, further decreasing the s-process production rate. These effects combine to produce a limiting metallicity ($[\mathrm{Fe/H}] \approx -3.5$) below which the `standard' s-process contribution becomes negligible \citep{Prantzos:1990}. There are, however, variations of the main s-process that can occur at low metallicity.  
 
One example is the rotationally-boosted s-process, where a spread in strontium production is predicted as a natural consequence \citep{pignatari:2008,frischknecht:2012,frischknecht:2016}. High initial rotation ($\vinicrit = 0.4$) in 15 to 40\,$M_\odot$ stars can drive rotational mixing that allows for large amounts of $^{14}$N to be produced in the hydrogen-burning shell \citep{Meynet:2002a,Meynet:2002b,Hirschi:2007}. A significant amount of $^{14}$N is eventually engulfed into the helium-burning core, where it is transformed into $^{22}$Ne through the $^{14}\mathrm{N}(\alpha,\gamma)^{18}\mathrm{F}(\beta^+\nu)^{18}\mathrm{O}(\alpha,\gamma)^{22}\mathrm{Ne}$ reaction chain. As stated earlier, the increase in $^{22}\mathrm{Ne}$ favours s-process production because the main neutron source for the s-process at low metallicity is produced through the $^{22}\mathrm{Ne}(\alpha,\mathrm{n})^{25}\mathrm{Mg}$ reaction. At this point s-process nucleosynthesis is limited only by the number of seeds (Fe nuclei) available, as the amount of neutron poisons does not change with rotation. Indeed, rotation largely only acts through the $^{22}\mathrm{Ne}$ source, as changes to the size of the hydrogen- and helium-burning regions due to rotation have a relatively small effect on nucleosynthetic yields.

Rotation can increase s-process nucleosynthesis in both metal-poor and metal-rich stars, but the nucleosynthetic signature is more distinctive at lower metallicities. In massive stars the neutron source is less efficient than in low-mass asymptotic giant branch stars, which shifts the element production to the first peak (e.g., Sr) instead of later peaks (e.g., Ba). The s-process elements produced are then lost to the interstellar medium, either during substantial mass loss in winds in the final stages of stellar evolution \citep[e.g.,][]{Banerjee:2019}, or after the massive star explodes in a core-collapse supernova. Stars more massive than 40\,$M_\odot$ are not considered to contribute to s-process nucleosynthesis because they are thought to collapse directly into black holes \citep{Woosley:2002,Heger:2003}.

We hypothesise that the Phoenix stream Sr abundances are consistent with enrichment from a massive fast-rotating metal-poor star\footnote{Or stars, but we find a single massive fast-rotating metal-poor star provides a sufficient description for the abundances.} that has undergone s-process nucleosynthesis, and enriched the surrounding interstellar medium with neutron-capture elements. In this scenario, since the Phoenix star cluster was likely low mass -- and not massive enough to retain ejecta -- we suggest that the massive fast-rotating star existed in a previous generation of star formation, before the cluster reached $[\mathrm{Fe/H}] \approx -2.7$. The nucleosynthetic products of that massive fast-rotating star were then likely inhomogenously mixed into the pre-Phoenix gas cloud, which would account for the variations in Sr abundances of individual stars in the present generation. For these reasons, we only consider massive fast-rotating metal-poor star models below this metallicity.

We compared the Phoenix stream abundances to nucleosynthetic yields for fast-rotating metal-poor ($\vinicrit = 0.4$; $[\mathrm{Fe/H}] = -3.8$) stars of 15 to 40\,$M_\odot$ \citep{frischknecht:2016}. The $\vinicrit$ ratio defines the initial rotational velocity relative to the critical breakup speed. We also considered models with $[\mathrm{Fe/H}] = -5.8$, and models that rotate up to $\vinicrit = 0.6$, but found them to be unreasonable fits to the data. We note that the yields provided in \citet{frischknecht:2016} are pre-explosive yields, only up to the end of oxygen burning. We assume that the total s-process yields are not strongly modified by core collapse supernovae \citep[e.g.,][]{Tur:2009}, and restrict our comparison to heavy elements \citep[$Z > 26$; however, see][]{Maeder:2015, Choplin:2019}. We show the expected yields in the top panel of Figure~\ref{fig:yield-predictions}, assuming that $\approx10^{-5}\,M_\odot$ of Fe is produced during core-collapse. If we increase the amount of Fe produced to exceed $10^{-4}\,M_\odot$ then we find poor fits to the [Ni/Fe] measurements. However, the relationship between the amount of Fe produced and the final abundance patterns primarily depend on how much of the s-process elements are deposited into the interstellar medium through stellar winds, and how much are released during core-collapse. As we expand upon below, the first-order parameters that control Sr production here are the rotation rate and the stellar mass.

In these massive fast-rotating metal-poor star models the rotation rate largely governs the abundance ratios of first (Sr) to second (Ba) peak s-process elements, as long as there are sufficient seeds available. The set of models available to us only include zero or high rotation ($\vinicrit = 0$ or $>0.4$), but we find a reasonable quantitative fit from the high rotation models. The non-rotating models (equivalent in all other parameters) do not produce any appreciable amounts of Sr or Ba, or Fe-peak elements. For this reason, we find that the non-rotating models cannot reproduce the abundance ratios we observe. It is reasonable to assume that a different initial rotation speed, for the same set of model masses and metallicities, could provide a better description for these data \citep{Limongi:2018,Rizzuti:2019}. 

The 15\,$M_\odot$ model provides the most consistent fit to the mean abundances, but we find that a 20--25\,$M_\odot$ model would also provide a reasonable explanation. All heavy elements measured (Co, Ni, Cu, Zn, Sr, Ba) contribute to placing these constraints. Similarly, with these models if we assume no additional Fe production then the models tend to over-produce [Zn/Fe] (and heavier elements). While we only have one measurement of [Zn/Fe] -- in the star with the highest [Sr/Fe] measurement -- the upper limits of [Zn/Fe] from other stars helps constrain the range of suitable models. All of these fast-rotating metal-poor star models are consistent with our observations in that they over-produce [Sr/Fe] relative to [Ba/Fe]. 
We note that the abundance comparisons we are making here explicitly ignore any nucleosynthesis contributions that occur between the enrichment from the massive fast-rotating star and the present day. 

The predicted nucleosynthetic yields from massive and fast-rotating metal-poor stars clearly provide a good description for the neutron-capture abundances we observe in the Phoenix stream. However, there is another potential issue with this scenario that is worth commenting on. An argument proposed in favour of fast-rotation in metal-poor stars is that they are more compact than their metal-rich equivalents, and thus metal-poor stars should rotate faster \citep{frischknecht:2016}. However, evidence suggests that the stellar multiplicity fraction increases (to $\approx 100$\%) as overall metallicity decreases, and as primary mass increases, implying that nearly all massive and metal-poor stars should be in binaries \citep[e.g.,][]{Badenes:2018}. In a binary system the rotation of both stars is constrained by the orbital period of the system, which can never be shorter than a few days. This sets a critical upper limit for the rotation speed of nearly all massive metal-poor stars, which is far less than the $v_\mathrm{ini}/v_\mathrm{crit} = 0.4$ ratios considered here. This might make fast-rotating massive and/or metal-poor stars a rare occurrence due to the increased multiplicity fraction for those kinds of systems, even though these models provide a good match to our observations.

We also considered predicted yields from metal-poor asymptotic giant branch (AGB) stars, the dominant source of the main s-process in the present day. The main s-process can still occur in metal-poor environments, but it is less efficient for reasons described earlier. The typical yields of most low-metallicity AGB stars produce more barium relative to strontium \citep[e.g.,][]{Fishlock:2014}, unlike what we see in the Phoenix stream. We compared the Phoenix stream abundances to yields from 102 metal-poor ($Z = 0.0001$) low- and intermediate-mass AGB star models \citep{Lugaro:2012, Karakas:2014} and find that only very few (intermediate-mass) models can produce significantly higher [Sr/H] than [Ba/H]. Even among those intermediate-mass AGB models, the observed [Ba/Fe] ratio is an order of magnitude smaller than those predicted by the models.  While low- and intermediate-mass AGB stars would be an appealing scenario, we conclude that this is an unlikely explanation for the neutron-capture abundances in the Phoenix stream.

However, the r-process could also qualitatively reproduce the abundances in the Phoenix stream. Neutrino-driven winds produce neutron-capture elements through the r-process, with net yields that tend to favour `lighter' elements like Sr and Ba, instead of Eu \citep{Arcones:2013, Wanajo:2013}. Magneto-rotational instabilities during a core-collapse supernova is another example where heavy elements are produced through the r-process, but favouring elements between $Z \sim 35$ to $55$. In the lower panel of Figure~\ref{fig:yield-predictions} we show yield predictions for core-collapse supernovae that are driven by magneto-rotational instabilities \citep{Nishimura:2017}. Multiple models are shown, representing models dominated by magnetically-driven jets (low $\hat{L}_\nu$: $\hat{L}_\nu \lesssim 0.4$), or intermediate models that are driven in part by magnetic fields and neutrino heating ($\hat{L}_\nu \approx 0.5$ to $0.75$). We also considered models that are purely dominated by neutrino heating ($\hat{L}_\nu \gtrsim 1.0$) but these models under-predicted Ba (relative to Sr) by orders of magnitude, and are not shown in Figure~\ref{fig:yield-predictions}. Depending on the properties of the driving mechanism, these models can qualitatively produce a high [Sr/Ba] abundance ratio, like that observed in the Phoenix stream. The magnetic-driven models (low $\hat{L}_\nu$) tend to be in better agreement with the Phoenix stream abundances, and Fe-peak elements (Co, Ni, Cu, Zn) are under-predicted by all of these models. This could conceivably be explained by additional supernovae that contribute to the interstellar medium before the Phoenix cluster formed, but overall we conclude that core collapse supernovae that are driven by magneto-rotational instabilities tend to be a poorer explanation for the Phoenix stream observations.

One final enrichment scenario that could explain the Phoenix stream abundances is an electron-capture supernova (ECSN). 
These explosions occur from low mass ($\lesssim10\,M_\odot$) super AGB stars that form an electron-degenerate oxygen-neon-magnesium core mass of around $\sim{}1.1~M_\odot$. Hydrogen and helium shell burning continuously increases the core mass, forcing the core to contract. As the central density rises, electron capture onto $^{24}$Mg is induced, reducing electron density and further accelerating core contraction. Eventually, electron capture on $^{20}$Ne can lead to an ignition that leads to an ECSN \citep{Nomoto:2017}. This is one of the candidate sites for the r-process, which can account for much of the heavy element nucleosynthesis at low metallicities \citep{Kobayashi:2020}. In particular, the galactic enrichment of Sr at low metallicities $[\mathrm{Fe/H}] \approx -3.8$ can be entirely explained by ECSN, without the need for a lighter elements primary process \citep[LEPP;][]{Cristallo:2015,Kobayashi:2020}. 
Detailed simulations of $8.8\,M_\odot$ and $9.6\,M_\odot$ progenitors show that Sr, Zr, and Zn are over-produced during ECSN \citep{Wanajo:2018}, with small amounts of Ni and Cu, and negligible quantities of Ba and Eu produced. This qualitatively matches the abundance pattern observed in the Phoenix stream, if we assume that the products of an ECSN are inhomogenously mixed into the pre-Phoenix gas cloud, just like the massive fast-rotating metal-poor star scenario. The occurrence of ECSN is not well known -- the fraction of CCSN that are ignited by electron-capture varies between $f_\mathrm{ECSN} = 0.01$ to $0.3$ \citep{Wanajo:2009} --  but ECSN are at least frequent enough to account for the galactic enrichment of Sr at low metallicity \citep{Kobayashi:2020}. For these reasons, a single\footnote{Or many.} ECSN is a plausible alternative for the neutron-capture abundances in the Phoenix stream. Unfortunately, given the limited abundance measurements and the lack of published ECSN yields, it is difficult to further constrain this scenario.

We do not consider neutron star mergers here, or neutrino winds during those mergers. There are considerations that the delay-time distribution of coalescence times may make neutron star mergers an unlikely contributor in very low metallicity environments \cite[e.g.,][]{Cote:2019}. It is also generally considered that neutron star mergers produce large amounts of heavy r-process elements, including Ba, and would not satisfy the low Ba scatter constraint. However, there is evidence that the kilonova associated with the GW170817 event appears to have produced a very high Sr/Eu ratio, about ten times higher than the solar r-process pattern \citep{Ji:2019}. If later observations and nucleosynthesis calculations demonstrate that neutron star merger yields are dominated by light r-process elements like Sr, then this may represent a plausible explanation for the Sr scatter in the Phoenix stream. But we consider this scenario unlikely given current models and observations.

\subsection{Phoenix-10, the lithium-rich giant star}

Phoenix-10 shows an over-abundance of lithium for a typical giant star, which is likely unrelated to the other chemical abundance patterns in the stream. Lithium is a fragile element that is difficult to produce in net quantities in standard single star evolution \citep[e.g.,][]{Iben:1967}. In most stars the surface lithium abundance decreases as they evolve off the main-sequence and experience first dredge up, where surface layers are mixed into deeper, hotter regions. As a consequence, most giant stars have very little lithium at their surface, as is the case for the other Phoenix stars. 
However, it is also known that about $\sim$1\% of red giants are classified as lithium-rich, ${A(\textrm{Li}) > 1.5}$\,dex. The incidence of lithium-rich giants increases with overall metallicity from $\approx 0.5\%$ at $[\mathrm{Fe/H}] = -2$ to ${\approx 1.5}$\% at $[\mathrm{Fe/H}] = +0.5$ \citep{Casey:2019}. At $[\textrm{Fe/H}] \sim -2.9$, Phoenix-10 is among the most metal-poor lithium-rich giants known \citep{Casey:2016}.

Lithium-rich giants have been discovered in the Milky Way disk, halo, and in open and globular star clusters. Most are discovered serendipitously \citep[for a literature compendium see][]{Casey:2016}, and currently about 4,000 lithium-rich giants are known \citep{Casey:2019,Gao:2019}. While many mechanisms have been proposed to explain lithium-rich giants, one that remains consistent with the observations across the entire giant branch is where tidal interactions in a binary system drive sufficient mixing to produce lithium internally and bring it up to the surface \citep{Casey:2019}. This would allow for lithium enrichment at any point along the red giant branch, and would naturally explain the high frequency ($80^{+7}_{-6}$\%) of lithium-rich giants that have helium-burning cores \citep{Casey:2019}. We have two epochs of radial velocity measurements for Phoenix-10, separated by about one month. The radial velocity difference between epochs is $\approx $1\,km\,s$^{-1}$ and within the joint uncertainties. If the enhanced lithium in Phoenix-10 were due to tidal interactions in a binary system then the expected radial velocity variation could be much smaller than our radial velocity precision ($\approx\,1~\mathrm{km~s}^{-1}$), as a large range of orbital periods and mass ratios are capable of driving sufficient tidal interactions. 

\section{Conclusions} 
\label{sec:conclusions}

We confirm that the Phoenix stream is consistent with being a disrupted globular cluster. From high-resolution spectra we find the mean metallicity is $\meanFeIIH$ and the low intrinsic metallicity spread ($\spreadFeIIH$), is consistent with zero. The large peak-to-peak range of abundance ratios in [Al/Fe] and [Na/Fe] are consistent with metal-poor globular clusters. 

The neutron-capture abundance ratios in the Phoenix stream are not seen in any other globular cluster. Low [Ba/Fe] abundance ratios, with no measurable spread in [Ba/Fe] across the stream, are accompanied by a large intrinsic scatter in [Sr/Fe]. We find that the rotationally-boosted s-process from a massive fast-rotating metal-poor star ($15\,M_\odot$, $\vinicrit = 0.4$, $[\mathrm{Fe/H}] = -3.8$) provides the best explanation for these abundances. Since the Phoenix progenitor system was likely low mass, it suggests that this fast-rotating metal-poor star enriched the interstellar medium at very high redshift, before the cluster formed.

While the Phoenix progenitor system would be the most metal-poor globular cluster known, there is every likelihood that many comparable systems existed and have since disrupted \citep{Larsen:2012,Beasley:2019,Krumholz:2019}. The relative isolation of Phoenix in the Galactic halo, with an orbital inclination of $60^\circ$ relative to the disk, and a pericenter of 13\,kpc \citep{Wan:2020}, has helped prevent the Phoenix stars becoming well-mixed with the bulk of the Milky Way stellar population and left them detectable as a stellar stream.

We have shown how the detailed abundances of ancient metal-poor stars can provide unique insight on nucleosynthesis events in the very early universe. Ultra faint dwarf galaxies, globular clusters, and ancient metal-poor stars in the Milky Way halo all offer the prospect for uncovering stars with unusual chemical abundance signatures \citep[see discussion in][]{Karlsson:2013}. However, the Phoenix stream represents an example of a potentially rich source for ancient stars with peculiar chemical abundances: identifying disrupted systems in phase space can reveal chemical records of nucleosynthesis events. These relics are rare, but each of them contribute to our knowledge of the very early universe.

\software{\texttt{Stan} \citep{Carpenter:2017,Salvatier:2016}, 
		  \texttt{ATLAS} \citep{Castelli:2004},
	      \texttt{MOOG} \citep{Sneden:1973,Sobeck:2011},
		  \texttt{SMHR} \citep{Casey:2014}.
}

\acknowledgements
We thank the anonymous referee for a timely and fastidious report which improved this paper.
We thank 
	Amanda Karakas (Monash) for providing yields of metal-poor low- and intermediate-mass AGB stars, 
	David Yong (ANU) for help in interpreting the yield comparisons with ROA\,276,
	and
	Kevin Schlaufman (JHU) for productive conversations.
This paper includes data gathered with the 6.5 meter Magellan Telescopes located at Las Campanas Observatory, Chile. 
A.~R.~C. is supported in part by the Australian Research Council through a Discovery Early Career Researcher Award (DE190100656). Parts of this research were supported by the Australian Research Council Centre of Excellence for All Sky Astrophysics in 3 Dimensions (ASTRO 3D), through project number CE170100013.

TSL is supported by NASA through Hubble Fellowship grant HF2-51439.001 awarded by the Space Telescope Science Institute, which is operated by the Association of Universities for Research in Astronomy, Inc., for NASA, under contract NAS5-26555.
TSL, APJ, and YYM are supported by NASA through Hubble Fellowship grant HST-HF2-51439.001, HST-HF2-51393.001, and HST-HF2-51441.001 respectively, awarded by the Space Telescope Science Institute, which is operated by the Association of Universities for Research in Astronomy, Inc., for NASA, under contract NAS5-26555.
SK was partially supported by NSF grant AST-1813881 and Heising-Simons foundation grant 2018-1030.
J.~D.~Simpson, S.~L.~M., and D.~B.~Z. acknowledge the support of the Australian Research Council (ARC) through Discovery Project grant DP180101791.  
G.~S.~Da~C. also acknowledges ARC support through Discovery Project grant DP150103294. 
J.~D.~Simon was supported in part by the National Science Foundation through grant AST-1714873. 
A.~B.~P. is supported by NSF grant AST-1813881.

This research has made use of the SIMBAD database, operated at CDS, Strasbourg, France \citep{Simbad}, and NASA's Astrophysics Data System Bibliographic Services.

This work presents results from the European Space Agency (ESA) space
mission Gaia. Gaia data are being processed by the Gaia Data
Processing and Analysis Consortium (DPAC). Funding for the DPAC is
provided by national institutions, in particular the institutions
participating in the Gaia MultiLateral Agreement (MLA). The Gaia
mission website is https://www.cosmos.esa.int/gaia. The Gaia archive
website is https://archives.esac.esa.int/gaia.

This project used public archival data from the Dark Energy Survey
(DES). Funding for the DES Projects has been provided by the
U.S. Department of Energy, the U.S. National Science Foundation, the
Ministry of Science and Education of Spain, the Science and Technology
Facilities Council of the United Kingdom, the Higher Education Funding
Council for England, the National Center for Supercomputing
Applications at the University of Illinois at Urbana-Champaign, the
Kavli Institute of Cosmological Physics at the University of Chicago,
the Center for Cosmology and Astro-Particle Physics at the Ohio State
University, the Mitchell Institute for Fundamental Physics and
Astronomy at Texas A\&M University, Financiadora de Estudos e
Projetos, Funda{\c c}{\~a}o Carlos Chagas Filho de Amparo {\`a}
Pesquisa do Estado do Rio de Janeiro, Conselho Nacional de
Desenvolvimento Cient{\'i}fico e Tecnol{\'o}gico and the
Minist{\'e}rio da Ci{\^e}ncia, Tecnologia e Inova{\c c}{\~a}o, the
Deutsche Forschungsgemeinschaft, and the Collaborating Institutions in
the Dark Energy Survey.  The Collaborating Institutions are Argonne
National Laboratory, the University of California at Santa Cruz, the
University of Cambridge, Centro de Investigaciones Energ{\'e}ticas,
Medioambientales y Tecnol{\'o}gicas-Madrid, the University of Chicago,
University College London, the DES-Brazil Consortium, the University
of Edinburgh, the Eidgen{\"o}ssische Technische Hochschule (ETH)
Z{\"u}rich, Fermi National Accelerator Laboratory, the University of
Illinois at Urbana-Champaign, the Institut de Ci{\`e}ncies de l'Espai
(IEEC/CSIC), the Institut de F{\'i}sica d'Altes Energies, Lawrence
Berkeley National Laboratory, the Ludwig-Maximilians Universit{\"a}t
M{\"u}nchen and the associated Excellence Cluster Universe, the
University of Michigan, the National Optical Astronomy Observatory,
the University of Nottingham, The Ohio State University, the OzDES
Membership Consortium, the University of Pennsylvania, the University
of Portsmouth, SLAC National Accelerator Laboratory, Stanford
University, the University of Sussex, and Texas A\&M University.
Based in part on observations at Cerro Tololo Inter-American
Observatory, National Optical Astronomy Observatory, which is operated
by the Association of Universities for Research in Astronomy (AURA)
under a cooperative agreement with the National Science Foundation.

The Legacy Surveys consist of three individual and complementary
projects: the Dark Energy Camera Legacy Survey (DECaLS; NOAO Proposal
ID \# 2014B-0404; PIs: David Schlegel and Arjun Dey), the
Beijing-Arizona Sky Survey (BASS; NOAO Proposal ID \# 2015A-0801; PIs:
Zhou Xu and Xiaohui Fan), and the Mayall z-band Legacy Survey (MzLS;
NOAO Proposal ID \# 2016A-0453; PI: Arjun Dey). DECaLS, BASS and MzLS
together include data obtained, respectively, at the Blanco telescope,
Cerro Tololo Inter-American Observatory, National Optical Astronomy
Observatory (NOAO); the Bok telescope, Steward Observatory, University
of Arizona; and the Mayall telescope, Kitt Peak National Observatory,
NOAO. The Legacy Surveys project is honored to be permitted to conduct
astronomical research on Iolkam Du'ag (Kitt Peak), a mountain with
particular significance to the Tohono O'odham Nation.

NOAO is operated by the Association of Universities for Research in
Astronomy (AURA) under a cooperative agreement with the National
Science Foundation.

The Legacy Survey team makes use of data products from the Near-Earth
Object Wide-field Infrared Survey Explorer (NEOWISE), which is a
project of the Jet Propulsion Laboratory/California Institute of
Technology. NEOWISE is funded by the National Aeronautics and Space
Administration.

The Legacy Surveys imaging of the DESI footprint is supported by the
Director, Office of Science, Office of High Energy Physics of the
U.S. Department of Energy under Contract No. DE-AC02-05CH1123, by the
National Energy Research Scientific Computing Center, a DOE Office of
Science User Facility under the same contract; and by the
U.S. National Science Foundation, Division of Astronomical Sciences
under Contract No. AST-0950945 to NOAO.

The Pan-STARRS1 Surveys (PS1) and the PS1 public science archive have been made possible through contributions by the Institute for Astronomy, the University of Hawaii, the Pan-STARRS Project Office, the Max-Planck Society and its participating institutes, the Max Planck Institute for Astronomy, Heidelberg and the Max Planck Institute for Extraterrestrial Physics, Garching, The Johns Hopkins University, Durham University, the University of Edinburgh, the Queen's University Belfast, the Harvard-Smithsonian Center for Astrophysics, the Las Cumbres Observatory Global Telescope Network Incorporated, the National Central University of Taiwan, the Space Telescope Science Institute, the National Aeronautics and Space Administration under Grant No. NNX08AR22G issued through the Planetary Science Division of the NASA Science Mission Directorate, the National Science Foundation Grant No. AST-1238877, the University of Maryland, Eotvos Lorand University (ELTE), the Los Alamos National Laboratory, and the Gordon and Betty Moore Foundation.

This manuscript has been authored by Fermi Research Alliance, LLC under Contract No. DE-AC02-07CH11359 with the U.S. Department of Energy, Office of Science, Office of High Energy Physics. The United States Government retains and the publisher, by accepting the article for publication, acknowledges that the United States Government retains a non-exclusive, paid-up, irrevocable, world-wide license to publish or reproduce the published form of this manuscript, or allow others to do so, for United States Government purposes.

\bibliography{ms}{}

\begin{thebibliography}{}
\expandafter\ifx\csname natexlab\endcsname\relax\def\natexlab#1{#1}\fi
\providecommand{\url}[1]{\href{#1}{#1}}
\providecommand{\dodoi}[1]{doi:~\href{http://doi.org/#1}{\nolinkurl{#1}}}
\providecommand{\doeprint}[1]{\href{http://ascl.net/#1}{\nolinkurl{http://ascl.net/#1}}}
\providecommand{\doarXiv}[1]{\href{https://arxiv.org/abs/#1}{\nolinkurl{https://arxiv.org/abs/#1}}}

\bibitem[{{Abohalima} \& {Frebel}(2018)}]{Abohalima:2018}
{Abohalima}, A., \& {Frebel}, A. 2018, \apjs, 238, 36,
  \dodoi{10.3847/1538-4365/aadfe9}

\bibitem[{{Andrievsky} {et~al.}(2011){Andrievsky}, {Spite}, {Korotin},
  {Fran{\c{c}}ois}, {Spite}, {Bonifacio}, {Cayrel}, \&
  {Hill}}]{Andrievsky:2011}
{Andrievsky}, S.~M., {Spite}, F., {Korotin}, S.~A., {et~al.} 2011, \aap, 530,
  A105, \dodoi{10.1051/0004-6361/201116591}

\bibitem[{{Aoki} {et~al.}(2009){Aoki}, {Arimoto}, {Sadakane}, {Tolstoy},
  {Battaglia}, {Jablonka}, {Shetrone}, {Letarte}, {Irwin}, {Hill}, {Francois},
  {Venn}, {Primas}, {Helmi}, {Kaufer}, {Tafelmeyer}, {Szeifert}, \&
  {Babusiaux}}]{Aoki:2009}
{Aoki}, W., {Arimoto}, N., {Sadakane}, K., {et~al.} 2009, \aap, 502, 569,
  \dodoi{10.1051/0004-6361/200911959}

\bibitem[{{Arcones} \& {Thielemann}(2013)}]{Arcones:2013}
{Arcones}, A., \& {Thielemann}, F.~K. 2013, Journal of Physics G Nuclear
  Physics, 40, 013201, \dodoi{10.1088/0954-3899/40/1/013201}

\bibitem[{{Badenes} {et~al.}(2018){Badenes}, {Mazzola}, {Thompson}, {Covey},
  {Freeman}, {Walker}, {Moe}, {Troup}, {Nidever}, {Allende Prieto}, {Andrews},
  {Barb{\'a}}, {Beers}, {Bovy}, {Carlberg}, {De Lee}, {Johnson}, {Lewis},
  {Majewski}, {Pinsonneault}, {Sobeck}, {Stassun}, {Stringfellow}, \&
  {Zasowski}}]{Badenes:2018}
{Badenes}, C., {Mazzola}, C., {Thompson}, T.~A., {et~al.} 2018, \apj, 854, 147,
  \dodoi{10.3847/1538-4357/aaa765}

\bibitem[{{Balbinot} {et~al.}(2016){Balbinot}, {Yanny}, {Li}, {Santiago},
  {Marshall}, {Finley}, {Pieres}, {Abbott}, {Abdalla}, {Allam},
  {Benoit-L{\'e}vy}, {Bernstein}, {Bertin}, {Brooks}, {Burke}, {Carnero
  Rosell}, {Carrasco Kind}, {Carretero}, {Cunha}, {da Costa}, {DePoy}, {Desai},
  {Diehl}, {Doel}, {Estrada}, {Flaugher}, {Frieman}, {Gerdes}, {Gruen},
  {Gruendl}, {Honscheid}, {James}, {Kuehn}, {Kuropatkin}, {Lahav}, {March},
  {Martini}, {Miquel}, {Nichol}, {Ogando}, {Romer}, {Sanchez}, {Schubnell},
  {Sevilla-Noarbe}, {Smith}, {Soares-Santos}, {Sobreira}, {Suchyta}, {Tarle},
  {Thomas}, {Tucker}, {Walker}, \& {DES Collaboration}}]{Balbinot:2016}
{Balbinot}, E., {Yanny}, B., {Li}, T.~S., {et~al.} 2016, \apj, 820, 58,
  \dodoi{10.3847/0004-637X/820/1/58}

\bibitem[{{Banerjee} {et~al.}(2019){Banerjee}, {Heger}, \&
  {Qian}}]{Banerjee:2019}
{Banerjee}, P., {Heger}, A., \& {Qian}, Y.-Z. 2019, arXiv e-prints,
  arXiv:1906.07335.
\newblock \doarXiv{1906.07335}

\bibitem[{{Barklem} {et~al.}(2005){Barklem}, {Christlieb}, {Beers}, {Hill},
  {Bessell}, {Holmberg}, {Marsteller}, {Rossi}, {Zickgraf}, \&
  {Reimers}}]{Barklem:2005}
{Barklem}, P.~S., {Christlieb}, N., {Beers}, T.~C., {et~al.} 2005, \aap, 439,
  129, \dodoi{10.1051/0004-6361:20052967}

\bibitem[{{Beasley} {et~al.}(2019){Beasley}, {Leaman}, {Gallart}, {Larsen},
  {Battaglia}, {Monelli}, \& {Pedreros}}]{Beasley:2019}
{Beasley}, M.~A., {Leaman}, R., {Gallart}, C., {et~al.} 2019, \mnras, 487,
  1986, \dodoi{10.1093/mnras/stz1349}

\bibitem[{{Belokurov} {et~al.}(2007){Belokurov}, {Zucker}, {Evans}, {Kleyna},
  {Koposov}, {Hodgkin}, {Irwin}, {Gilmore}, {Wilkinson}, {Fellhauer},
  {Bramich}, {Hewett}, {Vidrih}, {De Jong}, {Smith}, {Rix}, {Bell}, {Wyse},
  {Newberg}, {Mayeur}, {Yanny}, {Rockosi}, {Gnedin}, {Schneider}, {Beers},
  {Barentine}, {Brewington}, {Brinkmann}, {Harvanek}, {Kleinman}, {Krzesinski},
  {Long}, {Nitta}, \& {Snedden}}]{Belokurov:2007}
{Belokurov}, V., {Zucker}, D.~B., {Evans}, N.~W., {et~al.} 2007, \apj, 654,
  897, \dodoi{10.1086/509718}

\bibitem[{{Bernstein} {et~al.}(2003){Bernstein}, {Shectman}, {Gunnels},
  {Mochnacki}, \& {Athey}}]{Bernstein:2003}
{Bernstein}, R., {Shectman}, S.~A., {Gunnels}, S.~M., {Mochnacki}, S., \&
  {Athey}, A.~E. 2003, in Society of Photo-Optical Instrumentation Engineers
  (SPIE) Conference Series, Vol. 4841, Instrument Design and Performance for
  Optical/Infrared Ground-based Telescopes, ed. M.~{Iye} \& A.~F.~M.
  {Moorwood}, 1694--1704, \dodoi{10.1117/12.461502}

\bibitem[{{Bonaca} {et~al.}(2012){Bonaca}, {Geha}, \&
  {Kallivayalil}}]{Bonaca:2012}
{Bonaca}, A., {Geha}, M., \& {Kallivayalil}, N. 2012, \apjl, 760, L6,
  \dodoi{10.1088/2041-8205/760/1/L6}

\bibitem[{Carpenter {et~al.}(2017)Carpenter, Gelman, Hoffman, Lee, Goodrich,
  Betancourt, Brubaker, Guo, Li, \& Riddell}]{Carpenter:2017}
Carpenter, B., Gelman, A., Hoffman, M.~D., {et~al.} 2017, Journal of
  Statistical Software, 76, \dodoi{10.18637/jss.v076.i01}

\bibitem[{{Carrera} {et~al.}(2013){Carrera}, {Pancino}, {Gallart}, \& {del
  Pino}}]{Carrera:2013}
{Carrera}, R., {Pancino}, E., {Gallart}, C., \& {del Pino}, A. 2013, \mnras,
  434, 1681, \dodoi{10.1093/mnras/stt1126}

\bibitem[{{Carretta}(2019)}]{Carretta:2019}
{Carretta}, E. 2019, \aap, 624, A24, \dodoi{10.1051/0004-6361/201935110}

\bibitem[{{Carretta} {et~al.}(2009){Carretta}, {Bragaglia}, {Gratton},
  {Lucatello}, {Catanzaro}, {Leone}, {Bellazzini}, {Claudi}, {D'Orazi},
  {Momany}, {Ortolani}, {Pancino}, {Piotto}, {Recio-Blanco}, \&
  {Sabbi}}]{Carretta:2009}
{Carretta}, E., {Bragaglia}, A., {Gratton}, R.~G., {et~al.} 2009, \aap, 505,
  117, \dodoi{10.1051/0004-6361/200912096}

\bibitem[{{Casagrande} \& {VandenBerg}(2014)}]{Casagrande:2014}
{Casagrande}, L., \& {VandenBerg}, D.~A. 2014, \mnras, 444, 392,
  \dodoi{10.1093/mnras/stu1476}

\bibitem[{{Casey}(2014)}]{Casey:2014}
{Casey}, A.~R. 2014, PhD thesis, Australian National University

\bibitem[{{Casey} {et~al.}(2016){Casey}, {Ruchti}, {Masseron}, {Randich},
  {Gilmore}, {Lind}, {Kennedy}, {Koposov}, {Hourihane}, {Franciosini}, {Lewis},
  {Magrini}, {Morbidelli}, {Sacco}, {Worley}, {Feltzing}, {Jeffries},
  {Vallenari}, {Bensby}, {Bragaglia}, {Flaccomio}, {Francois}, {Korn},
  {Lanzafame}, {Pancino}, {Recio-Blanco}, {Smiljanic}, {Carraro}, {Costado},
  {Damiani}, {Donati}, {Frasca}, {Jofr{\'e}}, {Lardo}, {de Laverny}, {Monaco},
  {Prisinzano}, {Sbordone}, {Sousa}, {Tautvai{\v{s}}ien{\.{e}}}, {Zaggia},
  {Zwitter}, {Delgado Mena}, {Chorniy}, {Martell}, {Silva Aguirre}, {Miglio},
  {Chiappini}, {Montalban}, {Morel}, \& {Valentini}}]{Casey:2016}
{Casey}, A.~R., {Ruchti}, G., {Masseron}, T., {et~al.} 2016, \mnras, 461, 3336,
  \dodoi{10.1093/mnras/stw1512}

\bibitem[{{Casey} {et~al.}(2019){Casey}, {Ho}, {Ness}, {Hogg}, {Rix},
  {Angelou}, {Hekker}, {Tout}, {Lattanzio}, {Karakas}, {Woods}, {Price-Whelan},
  \& {Schlaufman}}]{Casey:2019}
{Casey}, A.~R., {Ho}, A. Y.~Q., {Ness}, M., {et~al.} 2019, \apj, 880, 125,
  \dodoi{10.3847/1538-4357/ab27bf}

\bibitem[{{Castelli} \& {Kurucz}(2003)}]{Castelli:2004}
{Castelli}, F., \& {Kurucz}, R.~L. 2003, in IAU Symposium, Vol. 210, Modelling
  of Stellar Atmospheres, ed. N.~{Piskunov}, W.~W. {Weiss}, \& D.~F. {Gray},
  A20

\bibitem[{{Choplin}(2019)}]{Choplin:2019}
{Choplin}, A. 2019, arXiv e-prints, arXiv:1901.10708.
\newblock \doarXiv{1901.10708}

\bibitem[{{Cohen} {et~al.}(2013){Cohen}, {Christlieb}, {Thompson}, {McWilliam},
  {Shectman}, {Reimers}, {Wisotzki}, \& {Kirby}}]{Cohen:2013}
{Cohen}, J.~G., {Christlieb}, N., {Thompson}, I., {et~al.} 2013, \apj, 778, 56,
  \dodoi{10.1088/0004-637X/778/1/56}

\bibitem[{{C{\^o}t{\'e}} {et~al.}(2019){C{\^o}t{\'e}}, {Eichler}, {Arcones},
  {Hansen}, {Simonetti}, {Frebel}, {Fryer}, {Pignatari}, {Reichert},
  {Belczynski}, \& {Matteucci}}]{Cote:2019}
{C{\^o}t{\'e}}, B., {Eichler}, M., {Arcones}, A., {et~al.} 2019, \apj, 875,
  106, \dodoi{10.3847/1538-4357/ab10db}

\bibitem[{{Cristallo} {et~al.}(2015){Cristallo}, {Abia}, {Straniero}, \&
  {Piersanti}}]{Cristallo:2015}
{Cristallo}, S., {Abia}, C., {Straniero}, O., \& {Piersanti}, L. 2015, \apj,
  801, 53, \dodoi{10.1088/0004-637X/801/1/53}

\bibitem[{{DES Collaboration} {et~al.}(2018){DES Collaboration}, {Abbott},
  {Abdalla}, {Allam}, {Amara}, {Annis}, {Asorey}, {Avila}, {Ballester},
  {Banerji}, {Barkhouse}, {Baruah}, {Baumer}, {Bechtol}, {Becker},
  {Benoit-L{\'e}vy}, {Bernstein}, {Bertin}, {Blazek}, {Bocquet}, {Brooks},
  {Brout}, {Buckley-Geer}, {Burke}, {Busti}, {Campisano}, {Cardiel-Sas},
  {Carnero Rosell}, {Carrasco Kind}, {Carretero}, {Castander}, {Cawthon},
  {Chang}, {Chen}, {Conselice}, {Costa}, {Crocce}, {Cunha}, {D'Andrea}, {da
  Costa}, {Das}, {Daues}, {Davis}, {Davis}, {De Vicente}, {DePoy}, {DeRose},
  {Desai}, {Diehl}, {Dietrich}, {Dodelson}, {Doel}, {Drlica-Wagner}, {Eifler},
  {Elliott}, {Evrard}, {Farahi}, {Fausti Neto}, {Fernandez}, {Finley},
  {Flaugher}, {Foley}, {Fosalba}, {Friedel}, {Frieman}, {Garc{\'\i}a-Bellido},
  {Gaztanaga}, {Gerdes}, {Giannantonio}, {Gill}, {Glazebrook}, {Goldstein},
  {Gower}, {Gruen}, {Gruendl}, {Gschwend}, {Gupta}, {Gutierrez}, {Hamilton},
  {Hartley}, {Hinton}, {Hislop}, {Hollowood}, {Honscheid}, {Hoyle}, {Huterer},
  {Jain}, {James}, {Jeltema}, {Johnson}, {Johnson}, {Kacprzak}, {Kent},
  {Khullar}, {Klein}, {Kovacs}, {Koziol}, {Krause}, {Kremin}, {Kron}, {Kuehn},
  {Kuhlmann}, {Kuropatkin}, {Lahav}, {Lasker}, {Li}, {Li}, {Liddle}, {Lima},
  {Lin}, {L{\'o}pez-Reyes}, {MacCrann}, {Maia}, {Maloney}, {Manera}, {March},
  {Marriner}, {Marshall}, {Martini}, {McClintock}, {McKay}, {McMahon},
  {Melchior}, {Menanteau}, {Miller}, {Miquel}, {Mohr}, {Morganson}, {Mould},
  {Neilsen}, {Nichol}, {Nogueira}, {Nord}, {Nugent}, {Nunes}, {Ogand o}, {Old},
  {Pace}, {Palmese}, {Paz-Chinch{\'o}n}, {Peiris}, {Percival}, {Petravick},
  {Plazas}, {Poh}, {Pond}, {Porredon}, {Pujol}, {Refregier}, {Reil}, {Ricker},
  {Rollins}, {Romer}, {Roodman}, {Rooney}, {Ross}, {Rykoff}, {Sako}, {Sanchez},
  {Sanchez}, {Santiago}, {Saro}, {Scarpine}, {Scolnic}, {Serrano},
  {Sevilla-Noarbe}, {Sheldon}, {Shipp}, {Silveira}, {Smith}, {Smith}, {Smith},
  {Soares-Santos}, {Sobreira}, {Song}, {Stebbins}, {Suchyta}, {Sullivan},
  {Swanson}, {Tarle}, {Thaler}, {Thomas}, {Thomas}, {Troxel}, {Tucker},
  {Vikram}, {Vivas}, {Walker}, {Wechsler}, {Weller}, {Wester}, {Wolf}, {Wu},
  {Yanny}, {Zenteno}, {Zhang}, {Zuntz}, {DES Collaboration}, {Juneau},
  {Fitzpatrick}, {Nikutta}, {Nidever}, {Olsen}, {Scott}, \& {NOAO Data
  Lab}}]{DESDR1}
{DES Collaboration}, {Abbott}, T.~M.~C., {Abdalla}, F.~B., {et~al.} 2018,
  \apjs, 239, 18, \dodoi{10.3847/1538-4365/aae9f0}

\bibitem[{{Dotter} {et~al.}(2008){Dotter}, {Chaboyer}, {Jevremovi{\'c}},
  {Kostov}, {Baron}, \& {Ferguson}}]{Dotter:2008}
{Dotter}, A., {Chaboyer}, B., {Jevremovi{\'c}}, D., {et~al.} 2008, \apjs, 178,
  89, \dodoi{10.1086/589654}

\bibitem[{{Erkal} {et~al.}(2016){Erkal}, {Sanders}, \&
  {Belokurov}}]{Erkal:2016}
{Erkal}, D., {Sanders}, J.~L., \& {Belokurov}, V. 2016, \mnras, 461, 1590,
  \dodoi{10.1093/mnras/stw1400}

\bibitem[{{Fishlock} {et~al.}(2014){Fishlock}, {Karakas}, {Lugaro}, \&
  {Yong}}]{Fishlock:2014}
{Fishlock}, C.~K., {Karakas}, A.~I., {Lugaro}, M., \& {Yong}, D. 2014, \apj,
  797, 44, \dodoi{10.1088/0004-637X/797/1/44}

\bibitem[{{Frebel} {et~al.}(2010){Frebel}, {Simon}, {Geha}, \&
  {Willman}}]{Frebel:2010}
{Frebel}, A., {Simon}, J.~D., {Geha}, M., \& {Willman}, B. 2010, \apj, 708,
  560, \dodoi{10.1088/0004-637X/708/1/560}

\bibitem[{{Frischknecht} {et~al.}(2012){Frischknecht}, {Hirschi}, \&
  {Thielemann}}]{frischknecht:2012}
{Frischknecht}, U., {Hirschi}, R., \& {Thielemann}, F.~K. 2012, \aap, 538, L2,
  \dodoi{10.1051/0004-6361/201117794}

\bibitem[{{Frischknecht} {et~al.}(2016){Frischknecht}, {Hirschi}, {Pignatari},
  {Maeder}, {Meynet}, {Chiappini}, {Thielemann}, {Rauscher}, {Georgy}, \&
  {Ekstr{\"o}m}}]{frischknecht:2016}
{Frischknecht}, U., {Hirschi}, R., {Pignatari}, M., {et~al.} 2016, \mnras, 456,
  1803, \dodoi{10.1093/mnras/stv2723}

\bibitem[{{Fulbright}(2000)}]{Fulbright:2000}
{Fulbright}, J.~P. 2000, \aj, 120, 1841, \dodoi{10.1086/301548}

\bibitem[{{Gaia Collaboration} {et~al.}(2018){Gaia Collaboration}, {Brown},
  {Vallenari}, {Prusti}, {de Bruijne}, {Babusiaux}, {Bailer-Jones}, {Biermann},
  {Evans}, {Eyer}, {Jansen}, {Jordi}, {Klioner}, {Lammers}, {Lindegren},
  {Luri}, {Mignard}, {Panem}, {Pourbaix}, {Randich}, {Sartoretti}, {Siddiqui},
  {Soubiran}, {van Leeuwen}, {Walton}, {Arenou}, {Bastian}, {Cropper},
  {Drimmel}, {Katz}, {Lattanzi}, {Bakker}, {Cacciari}, {Casta{\~n}eda},
  {Chaoul}, {Cheek}, {De Angeli}, {Fabricius}, {Guerra}, {Holl}, {Masana},
  {Messineo}, {Mowlavi}, {Nienartowicz}, {Panuzzo}, {Portell}, {Riello},
  {Seabroke}, {Tanga}, {Th{\'e}venin}, {Gracia-Abril}, {Comoretto},
  {Garcia-Reinaldos}, {Teyssier}, {Altmann}, {Andrae}, {Audard},
  {Bellas-Velidis}, {Benson}, {Berthier}, {Blomme}, {Burgess}, {Busso},
  {Carry}, {Cellino}, {Clementini}, {Clotet}, {Creevey}, {Davidson}, {De
  Ridder}, {Delchambre}, {Dell'Oro}, {Ducourant},
  {Fern{\'a}ndez-Hern{\'a}ndez}, {Fouesneau}, {Fr{\'e}mat}, {Galluccio},
  {Garc{\'\i}a-Torres}, {Gonz{\'a}lez-N{\'u}{\~n}ez}, {Gonz{\'a}lez-Vidal},
  {Gosset}, {Guy}, {Halbwachs}, {Hambly}, {Harrison}, {Hern{\'a}ndez},
  {Hestroffer}, {Hodgkin}, {Hutton}, {Jasniewicz}, {Jean-Antoine-Piccolo},
  {Jordan}, {Korn}, {Krone-Martins}, {Lanzafame}, {Lebzelter}, {L{\"o}ffler},
  {Manteiga}, {Marrese}, {Mart{\'\i}n-Fleitas}, {Moitinho}, {Mora}, {Muinonen},
  {Osinde}, {Pancino}, {Pauwels}, {Petit}, {Recio-Blanco}, {Richards},
  {Rimoldini}, {Robin}, {Sarro}, {Siopis}, {Smith}, {Sozzetti}, {S{\"u}veges},
  {Torra}, {van Reeven}, {Abbas}, {Abreu Aramburu}, {Accart}, {Aerts},
  {Altavilla}, {{\'A}lvarez}, {Alvarez}, {Alves}, {Anderson}, {Andrei},
  {Anglada Varela}, {Antiche}, {Antoja}, {Arcay}, {Astraatmadja}, {Bach},
  {Baker}, {Balaguer-N{\'u}{\~n}ez}, {Balm}, {Barache}, {Barata}, {Barbato},
  {Barblan}, {Barklem}, {Barrado}, {Barros}, {Barstow}, {Bartholom{\'e}
  Mu{\~n}oz}, {Bassilana}, {Becciani}, {Bellazzini}, {Berihuete}, {Bertone},
  {Bianchi}, {Bienaym{\'e}}, {Blanco-Cuaresma}, {Boch}, {Boeche}, {Bombrun},
  {Borrachero}, {Bossini}, {Bouquillon}, {Bourda}, {Bragaglia}, {Bramante},
  {Breddels}, {Bressan}, {Brouillet}, {Br{\"u}semeister}, {Brugaletta},
  {Bucciarelli}, {Burlacu}, {Busonero}, {Butkevich}, {Buzzi}, {Caffau},
  {Cancelliere}, {Cannizzaro}, {Cantat-Gaudin}, {Carballo}, {Carlucci},
  {Carrasco}, {Casamiquela}, {Castellani}, {Castro-Ginard}, {Charlot},
  {Chemin}, {Chiavassa}, {Cocozza}, {Costigan}, {Cowell}, {Crifo}, {Crosta},
  {Crowley}, {Cuypers}, {Dafonte}, {Damerdji}, {Dapergolas}, {David}, {David},
  {de Laverny}, {De Luise}, {De March}, {de Martino}, {de Souza}, {de Torres},
  {Debosscher}, {del Pozo}, {Delbo}, {Delgado}, {Delgado}, {Di Matteo},
  {Diakite}, {Diener}, {Distefano}, {Dolding}, {Drazinos}, {Dur{\'a}n},
  {Edvardsson}, {Enke}, {Eriksson}, {Esquej}, {Eynard Bontemps}, {Fabre},
  {Fabrizio}, {Faigler}, {Falc{\~a}o}, {Farr{\`a}s Casas}, {Federici},
  {Fedorets}, {Fernique}, {Figueras}, {Filippi}, {Findeisen}, {Fonti},
  {Fraile}, {Fraser}, {Fr{\'e}zouls}, {Gai}, {Galleti}, {Garabato},
  {Garc{\'\i}a-Sedano}, {Garofalo}, {Garralda}, {Gavel}, {Gavras}, {Gerssen},
  {Geyer}, {Giacobbe}, {Gilmore}, {Girona}, {Giuffrida}, {Glass}, {Gomes},
  {Granvik}, {Gueguen}, {Guerrier}, {Guiraud}, {Guti{\'e}rrez-S{\'a}nchez},
  {Haigron}, {Hatzidimitriou}, {Hauser}, {Haywood}, {Heiter}, {Helmi}, {Heu},
  {Hilger}, {Hobbs}, {Hofmann}, {Holland}, {Huckle}, {Hypki}, {Icardi},
  {Jan{\ss}en}, {Jevardat de Fombelle}, {Jonker}, {Juh{\'a}sz}, {Julbe},
  {Karampelas}, {Kewley}, {Klar}, {Kochoska}, {Kohley}, {Kolenberg},
  {Kontizas}, {Kontizas}, {Koposov}, {Kordopatis}, {Kostrzewa-Rutkowska},
  {Koubsky}, {Lambert}, {Lanza}, {Lasne}, {Lavigne}, {Le Fustec}, {Le
  Poncin-Lafitte}, {Lebreton}, {Leccia}, {Leclerc}, {Lecoeur-Taibi},
  {Lenhardt}, {Leroux}, {Liao}, {Licata}, {Lindstr{\o}m}, {Lister}, {Livanou},
  {Lobel}, {L{\'o}pez}, {Managau}, {Mann}, {Mantelet}, {Marchal}, {Marchant},
  {Marconi}, {Marinoni}, {Marschalk{\'o}}, {Marshall}, {Martino}, {Marton},
  {Mary}, {Massari}, {Matijevi{\v{c}}}, {Mazeh}, {McMillan}, {Messina},
  {Michalik}, {Millar}, {Molina}, {Molinaro}, {Moln{\'a}r}, {Montegriffo},
  {Mor}, {Morbidelli}, {Morel}, {Morris}, {Mulone}, {Muraveva}, {Musella},
  {Nelemans}, {Nicastro}, {Noval}, {O'Mullane}, {Ord{\'e}novic},
  {Ord{\'o}{\~n}ez-Blanco}, {Osborne}, {Pagani}, {Pagano}, {Pailler},
  {Palacin}, {Palaversa}, {Panahi}, {Pawlak}, {Piersimoni}, {Pineau}, {Plachy},
  {Plum}, {Poggio}, {Poujoulet}, {Pr{\v{s}}a}, {Pulone}, {Racero}, {Ragaini},
  {Rambaux}, {Ramos-Lerate}, {Regibo}, {Reyl{\'e}}, {Riclet}, {Ripepi}, {Riva},
  {Rivard}, {Rixon}, {Roegiers}, {Roelens}, {Romero-G{\'o}mez}, {Rowell},
  {Royer}, {Ruiz-Dern}, {Sadowski}, {Sagrist{\`a} Sell{\'e}s}, {Sahlmann},
  {Salgado}, {Salguero}, {Sanna}, {Santana-Ros}, {Sarasso}, {Savietto},
  {Schultheis}, {Sciacca}, {Segol}, {Segovia}, {S{\'e}gransan}, {Shih},
  {Siltala}, {Silva}, {Smart}, {Smith}, {Solano}, {Solitro}, {Sordo}, {Soria
  Nieto}, {Souchay}, {Spagna}, {Spoto}, {Stampa}, {Steele},
  {Steidelm{\"u}ller}, {Stephenson}, {Stoev}, {Suess}, {Surdej}, {Szabados},
  {Szegedi-Elek}, {Tapiador}, {Taris}, {Tauran}, {Taylor}, {Teixeira},
  {Terrett}, {Teyssand ier}, {Thuillot}, {Titarenko}, {Torra Clotet}, {Turon},
  {Ulla}, {Utrilla}, {Uzzi}, {Vaillant}, {Valentini}, {Valette}, {van Elteren},
  {Van Hemelryck}, {van Leeuwen}, {Vaschetto}, {Vecchiato}, {Veljanoski},
  {Viala}, {Vicente}, {Vogt}, {von Essen}, {Voss}, {Votruba}, {Voutsinas},
  {Walmsley}, {Weiler}, {Wertz}, {Wevers}, {Wyrzykowski}, {Yoldas},
  {{\v{Z}}erjal}, {Ziaeepour}, {Zorec}, {Zschocke}, {Zucker}, {Zurbach}, \&
  {Zwitter}}]{Gaia:2018}
{Gaia Collaboration}, {Brown}, A.~G.~A., {Vallenari}, A., {et~al.} 2018, \aap,
  616, A1, \dodoi{10.1051/0004-6361/201833051}

\bibitem[{{Gao} {et~al.}(2019){Gao}, {Shi}, {Yan}, {Yan}, {Xiang}, {Zhou},
  {Li}, \& {Zhao}}]{Gao:2019}
{Gao}, Q., {Shi}, J.-R., {Yan}, H.-L., {et~al.} 2019, \apjs, 245, 33,
  \dodoi{10.3847/1538-4365/ab505c}

\bibitem[{{Hansen} {et~al.}(2013){Hansen}, {Bergemann}, {Cescutti},
  {Fran{\c{c}}ois}, {Arcones}, {Karakas}, {Lind}, \& {Chiappini}}]{Hansen:2013}
{Hansen}, C.~J., {Bergemann}, M., {Cescutti}, G., {et~al.} 2013, \aap, 551,
  A57, \dodoi{10.1051/0004-6361/201220584}

\bibitem[{{Heger} {et~al.}(2003){Heger}, {Fryer}, {Woosley}, {Langer}, \&
  {Hartmann}}]{Heger:2003}
{Heger}, A., {Fryer}, C.~L., {Woosley}, S.~E., {Langer}, N., \& {Hartmann},
  D.~H. 2003, \apj, 591, 288, \dodoi{10.1086/375341}

\bibitem[{{Hirschi}(2007)}]{Hirschi:2007}
{Hirschi}, R. 2007, \aap, 461, 571, \dodoi{10.1051/0004-6361:20065356}

\bibitem[{{Ibata} {et~al.}(1994){Ibata}, {Gilmore}, \& {Irwin}}]{Ibata:1994}
{Ibata}, R.~A., {Gilmore}, G., \& {Irwin}, M.~J. 1994, \nat, 370, 194,
  \dodoi{10.1038/370194a0}

\bibitem[{{Iben}(1967)}]{Iben:1967}
{Iben}, Icko, J. 1967, \apj, 147, 624, \dodoi{10.1086/149040}

\bibitem[{{Jacobson} {et~al.}(2015){Jacobson}, {Keller}, {Frebel}, {Casey},
  {Asplund}, {Bessell}, {Da Costa}, {Lind}, {Marino}, {Norris}, {Pe{\~n}a},
  {Schmidt}, {Tisserand}, {Walsh}, {Yong}, \& {Yu}}]{Jacobson:2015}
{Jacobson}, H.~R., {Keller}, S., {Frebel}, A., {et~al.} 2015, \apj, 807, 171,
  \dodoi{10.1088/0004-637X/807/2/171}

\bibitem[{{Ji} {et~al.}(2019){Ji}, {Simon}, {Frebel}, {Venn}, \&
  {Hansen}}]{Ji:2019}
{Ji}, A.~P., {Simon}, J.~D., {Frebel}, A., {Venn}, K.~A., \& {Hansen}, T.~T.
  2019, \apj, 870, 83, \dodoi{10.3847/1538-4357/aaf3bb}

\bibitem[{{Ji} {et~al.}(2020){Ji}, {Li}, {Hansen}, {Casey}, {Koposov}, {Pace},
  {Mackey}, {Lewis}, {Simpson}, {Bland-Hawthorn}, {Cullinane}, {Da Costa},
  {Hattori}, {Martell}, {Kuehn}, {Erkal}, {Shipp}, {Wan}, \&
  {Zucker}}]{Ji:2020}
{Ji}, A.~P., {Li}, T.~S., {Hansen}, T.~T., {et~al.} 2020, \aj, 160, 181,
  \dodoi{10.3847/1538-3881/abacb6}

\bibitem[{{Karakas} \& {Lattanzio}(2014)}]{Karakas:2014}
{Karakas}, A.~I., \& {Lattanzio}, J.~C. 2014, \pasa, 31, e030,
  \dodoi{10.1017/pasa.2014.21}

\bibitem[{{Karlsson} {et~al.}(2013){Karlsson}, {Bromm}, \&
  {Bland-Hawthorn}}]{Karlsson:2013}
{Karlsson}, T., {Bromm}, V., \& {Bland-Hawthorn}, J. 2013, Reviews of Modern
  Physics, 85, 809, \dodoi{10.1103/RevModPhys.85.809}

\bibitem[{{Kelson}(2003)}]{Kelson:2003}
{Kelson}, D.~D. 2003, \pasp, 115, 688, \dodoi{10.1086/375502}

\bibitem[{{Kobayashi} {et~al.}(2020){Kobayashi}, {Karakas}, \&
  {Lugaro}}]{Kobayashi:2020}
{Kobayashi}, C., {Karakas}, A.~I., \& {Lugaro}, M. 2020, \apj, 900, 179,
  \dodoi{10.3847/1538-4357/abae65}

\bibitem[{{Koppelman} {et~al.}(2019){Koppelman}, {Helmi}, {Massari},
  {Roelenga}, \& {Bastian}}]{Koppelman:2019}
{Koppelman}, H.~H., {Helmi}, A., {Massari}, D., {Roelenga}, S., \& {Bastian},
  U. 2019, \aap, 625, A5, \dodoi{10.1051/0004-6361/201834769}

\bibitem[{{Krumholz} {et~al.}(2019){Krumholz}, {McKee}, \&
  {Bland-Hawthorn}}]{Krumholz:2019}
{Krumholz}, M.~R., {McKee}, C.~F., \& {Bland-Hawthorn}, J. 2019, \araa, 57,
  227, \dodoi{10.1146/annurev-astro-091918-104430}

\bibitem[{{Kuzma} {et~al.}(2015){Kuzma}, {Da Costa}, {Keller}, \&
  {Maunder}}]{Kuzma:2015}
{Kuzma}, P.~B., {Da Costa}, G.~S., {Keller}, S.~C., \& {Maunder}, E. 2015,
  \mnras, 446, 3297, \dodoi{10.1093/mnras/stu2343}

\bibitem[{{Larsen} {et~al.}(2012){Larsen}, {Brodie}, \&
  {Strader}}]{Larsen:2012}
{Larsen}, S.~S., {Brodie}, J.~P., \& {Strader}, J. 2012, \aap, 546, A53,
  \dodoi{10.1051/0004-6361/201219895}

\bibitem[{{Li} {et~al.}(2019{\natexlab{a}}){Li}, {Koposov}, {Zucker}, {Lewis},
  {Kuehn}, {Simpson}, {Ji}, {Shipp}, {Mao}, {Geha}, {Pace}, {Mackey}, {Allam},
  {Tucker}, {Da Costa}, {Erkal}, {Simon}, {Mould}, {Martell}, {Wan}, {De
  Silva}, {Bechtol}, {Balbinot}, {Belokurov}, {Bland-Hawthorn}, {Casey},
  {Cullinane}, {Drlica-Wagner}, {Sharma}, {Vivas}, {Wechsler}, {Yanny}, \& {S5
  Collaboration}}]{Li:2019}
{Li}, T.~S., {Koposov}, S.~E., {Zucker}, D.~B., {et~al.} 2019{\natexlab{a}},
  \mnras, 490, 3508, \dodoi{10.1093/mnras/stz2731}

\bibitem[{{Li} {et~al.}(2019{\natexlab{b}}){Li}, {Koposov}, {Zucker}, {Lewis},
  {Kuehn}, {Simpson}, {Ji}, {Shipp}, {Mao}, {Geha}, {Pace}, {Mackey}, {Allam},
  {Tucker}, {Da Costa}, {Erkal}, {Simon}, {Mould}, {Martell}, {Wan}, {De
  Silva}, {Bechtol}, {Balbinot}, {Belokurov}, {Bland-Hawthorn}, {Casey},
  {Cullinane}, {Drlica-Wagner}, {Sharma}, {Vivas}, {Wechsler}, {Yanny}, \& {S5
  Collaboration}}]{Li:2020}
---. 2019{\natexlab{b}}, \mnras, 490, 3508, \dodoi{10.1093/mnras/stz2731}

\bibitem[{{Limongi} \& {Chieffi}(2018)}]{Limongi:2018}
{Limongi}, M., \& {Chieffi}, A. 2018, \apjs, 237, 13,
  \dodoi{10.3847/1538-4365/aacb24}

\bibitem[{{Lugaro} {et~al.}(2012){Lugaro}, {Karakas}, {Stancliffe}, \&
  {Rijs}}]{Lugaro:2012}
{Lugaro}, M., {Karakas}, A.~I., {Stancliffe}, R.~J., \& {Rijs}, C. 2012, \apj,
  747, 2, \dodoi{10.1088/0004-637X/747/1/2}

\bibitem[{{Maeder} {et~al.}(2015){Maeder}, {Meynet}, \&
  {Chiappini}}]{Maeder:2015}
{Maeder}, A., {Meynet}, G., \& {Chiappini}, C. 2015, \aap, 576, A56,
  \dodoi{10.1051/0004-6361/201424153}

\bibitem[{{Marino} {et~al.}(2011){Marino}, {Milone}, {Piotto}, {Villanova},
  {Gratton}, {D'Antona}, {Anderson}, {Bedin}, {Bellini}, {Cassisi}, {Geisler},
  {Renzini}, \& {Zoccali}}]{Marino:2011}
{Marino}, A.~F., {Milone}, A.~P., {Piotto}, G., {et~al.} 2011, \apj, 731, 64,
  \dodoi{10.1088/0004-637X/731/1/64}

\bibitem[{{Martin} {et~al.}(2018){Martin}, {Amy}, {Newberg}, {Shelton},
  {Carlin}, {Beers}, {Denissenkov}, \& {Willett}}]{Martin:2019}
{Martin}, C., {Amy}, P.~M., {Newberg}, H.~J., {et~al.} 2018, \mnras, 477, 2419,
  \dodoi{10.1093/mnras/sty608}

\bibitem[{{Meynet} \& {Maeder}(2002{\natexlab{a}})}]{Meynet:2002a}
{Meynet}, G., \& {Maeder}, A. 2002{\natexlab{a}}, \aap, 390, 561,
  \dodoi{10.1051/0004-6361:20020755}

\bibitem[{{Meynet} \& {Maeder}(2002{\natexlab{b}})}]{Meynet:2002b}
---. 2002{\natexlab{b}}, \aap, 381, L25, \dodoi{10.1051/0004-6361:20011554}

\bibitem[{{Morganson} {et~al.}(2018){Morganson}, {Gruendl}, {Menanteau},
  {Carrasco Kind}, {Chen}, {Daues}, {Drlica-Wagner}, {Friedel}, {Gower},
  {Johnson}, {Johnson}, {Kessler}, {Paz-Chinch{\'o}n}, {Petravick}, {Pond},
  {Yanny}, {Allam}, {Armstrong}, {Barkhouse}, {Bechtol}, {Benoit-L{\'e}vy},
  {Bernstein}, {Bertin}, {Buckley-Geer}, {Covarrubias}, {Desai}, {Diehl},
  {Goldstein}, {Gruen}, {Li}, {Lin}, {Marriner}, {Mohr}, {Neilsen}, {Ngeow},
  {Paech}, {Rykoff}, {Sako}, {Sevilla-Noarbe}, {Sheldon}, {Sobreira}, {Tucker},
  {Wester}, \& {DES Collaboration}}]{Morganson:2018}
{Morganson}, E., {Gruendl}, R.~A., {Menanteau}, F., {et~al.} 2018, \pasp, 130,
  074501, \dodoi{10.1088/1538-3873/aab4ef}

\bibitem[{{Nataf} {et~al.}(2019){Nataf}, {Wyse}, {Schiavon}, {Ting}, {Minniti},
  {Cohen}, {Fern{\'a}ndez-Trincado}, {Geisler}, {Nitschelm}, \&
  {Frinchaboy}}]{Nataf:2019}
{Nataf}, D.~M., {Wyse}, R. F.~G., {Schiavon}, R.~P., {et~al.} 2019, \aj, 158,
  14, \dodoi{10.3847/1538-3881/ab1a27}

\bibitem[{{Nishimura} {et~al.}(2017){Nishimura}, {Sawai}, {Takiwaki}, {Yamada},
  \& {Thielemann}}]{Nishimura:2017}
{Nishimura}, N., {Sawai}, H., {Takiwaki}, T., {Yamada}, S., \& {Thielemann},
  F.~K. 2017, \apjl, 836, L21, \dodoi{10.3847/2041-8213/aa5dee}

\bibitem[{Nomoto \& Leung(2017)}]{Nomoto:2017}
Nomoto, K., \& Leung, S.-C. 2017, Electron Capture Supernovae from Super
  Asymptotic Giant Branch Stars, ed. A.~W. Alsabti \& P.~Murdin (Cham: Springer
  International Publishing), 483--512, \dodoi{10.1007/978-3-319-21846-5_118}

\bibitem[{{Nordlander} \& {Lind}(2017)}]{Nordlander:2017}
{Nordlander}, T., \& {Lind}, K. 2017, \aap, 607, A75,
  \dodoi{10.1051/0004-6361/201730427}

\bibitem[{{Pancino}(2018)}]{Pancino:2018}
{Pancino}, E. 2018, \aap, 614, A80, \dodoi{10.1051/0004-6361/201732351}

\bibitem[{{Pignatari} {et~al.}(2008){Pignatari}, {Gallino}, {Meynet},
  {Hirschi}, {Herwig}, \& {Wiescher}}]{pignatari:2008}
{Pignatari}, M., {Gallino}, R., {Meynet}, G., {et~al.} 2008, \apjl, 687, L95,
  \dodoi{10.1086/593350}

\bibitem[{{Prantzos} {et~al.}(1990){Prantzos}, {Hashimoto}, \&
  {Nomoto}}]{Prantzos:1990}
{Prantzos}, N., {Hashimoto}, M., \& {Nomoto}, K. 1990, \aap, 234, 211

\bibitem[{{Rizzuti} {et~al.}(2019){Rizzuti}, {Cescutti}, {Matteucci},
  {Chieffi}, {Hirschi}, \& {Limongi}}]{Rizzuti:2019}
{Rizzuti}, F., {Cescutti}, G., {Matteucci}, F., {et~al.} 2019, \mnras, 489,
  5244, \dodoi{10.1093/mnras/stz2505}

\bibitem[{{Roederer} {et~al.}(2014){Roederer}, {Preston}, {Thompson},
  {Shectman}, {Sneden}, {Burley}, \& {Kelson}}]{Roederer:2014c}
{Roederer}, I.~U., {Preston}, G.~W., {Thompson}, I.~B., {et~al.} 2014, \aj,
  147, 136, \dodoi{10.1088/0004-6256/147/6/136}

\bibitem[{Salvatier {et~al.}(2016)Salvatier, Wiecki, \&
  Fonnesbeck}]{Salvatier:2016}
Salvatier, J., Wiecki, T.~V., \& Fonnesbeck, C. 2016, {PeerJ} Computer Science,
  2, e55, \dodoi{10.7717/peerj-cs.55}

\bibitem[{{Schlegel} {et~al.}(1998){Schlegel}, {Finkbeiner}, \&
  {Davis}}]{Schlegel:1998}
{Schlegel}, D.~J., {Finkbeiner}, D.~P., \& {Davis}, M. 1998, \apj, 500, 525,
  \dodoi{10.1086/305772}

\bibitem[{{Shipp} {et~al.}(2018){Shipp}, {Drlica-Wagner}, {Balbinot},
  {Ferguson}, {Erkal}, {Li}, {Bechtol}, {Belokurov}, {Buncher}, {Carollo},
  {Carrasco Kind}, {Kuehn}, {Marshall}, {Pace}, {Rykoff}, {Sevilla-Noarbe},
  {Sheldon}, {Strigari}, {Vivas}, {Yanny}, {Zenteno}, {Abbott}, {Abdalla},
  {Allam}, {Avila}, {Bertin}, {Brooks}, {Burke}, {Carretero}, {Castander},
  {Cawthon}, {Crocce}, {Cunha}, {D'Andrea}, {da Costa}, {Davis}, {De Vicente},
  {Desai}, {Diehl}, {Doel}, {Evrard}, {Flaugher}, {Fosalba}, {Frieman},
  {Garc{\'\i}a-Bellido}, {Gaztanaga}, {Gerdes}, {Gruen}, {Gruendl}, {Gschwend},
  {Gutierrez}, {Hartley}, {Honscheid}, {Hoyle}, {James}, {Johnson}, {Krause},
  {Kuropatkin}, {Lahav}, {Lin}, {Maia}, {March}, {Martini}, {Menanteau},
  {Miller}, {Miquel}, {Nichol}, {Plazas}, {Romer}, {Sako}, {Sanchez},
  {Santiago}, {Scarpine}, {Schindler}, {Schubnell}, {Smith}, {Smith},
  {Sobreira}, {Suchyta}, {Swanson}, {Tarle}, {Thomas}, {Tucker}, {Walker},
  {Wechsler}, \& {DES Collaboration}}]{Shipp:2018}
{Shipp}, N., {Drlica-Wagner}, A., {Balbinot}, E., {et~al.} 2018, \apj, 862,
  114, \dodoi{10.3847/1538-4357/aacdab}

\bibitem[{{Shipp} {et~al.}(2019){Shipp}, {Li}, {Pace}, {Erkal},
  {Drlica-Wagner}, {Yanny}, {Belokurov}, {Wester}, {Koposov}, {Kuehn}, {Lewis},
  {Simpson}, {Wan}, {Zucker}, {Martell}, {Wang}, \& {S5
  Collaboration}}]{Shipp:2019}
{Shipp}, N., {Li}, T.~S., {Pace}, A.~B., {et~al.} 2019, \apj, 885, 3,
  \dodoi{10.3847/1538-4357/ab44bf}

\bibitem[{{Sneden}(1973)}]{Sneden:1973}
{Sneden}, C.~A. 1973, PhD thesis, THE UNIVERSITY OF TEXAS AT AUSTIN.

\bibitem[{{Sobeck} {et~al.}(2011){Sobeck}, {Kraft}, {Sneden}, {Preston},
  {Cowan}, {Smith}, {Thompson}, {Shectman}, \& {Burley}}]{Sobeck:2011}
{Sobeck}, J.~S., {Kraft}, R.~P., {Sneden}, C., {et~al.} 2011, \aj, 141, 175,
  \dodoi{10.1088/0004-6256/141/6/175}

\bibitem[{{Stanford} {et~al.}(2010){Stanford}, {Da Costa}, \&
  {Norris}}]{Stanford:2010}
{Stanford}, L.~M., {Da Costa}, G.~S., \& {Norris}, J.~E. 2010, \apj, 714, 1001,
  \dodoi{10.1088/0004-637X/714/2/1001}

\bibitem[{{Tur} {et~al.}(2009){Tur}, {Heger}, \& {Austin}}]{Tur:2009}
{Tur}, C., {Heger}, A., \& {Austin}, S.~M. 2009, \apj, 702, 1068,
  \dodoi{10.1088/0004-637X/702/2/1068}

\bibitem[{{Wan} {et~al.}(2020){Wan}, {Lewis}, {Li}, {Simpson}, {Martell},
  {Zucker}, {Mould}, {Erkal}, {Pace}, {Mackey}, {Ji}, {Koposov}, {Kuehn},
  {Shipp}, {Balbinot}, {Bland-Hawthorn}, {Casey}, {Da Costa}, {Kafle},
  {Sharma}, \& {De Silva}}]{Wan:2020}
{Wan}, Z., {Lewis}, G.~F., {Li}, T.~S., {et~al.} 2020, \nat, 583, 768,
  \dodoi{10.1038/s41586-020-2483-6}

\bibitem[{{Wanajo}(2013)}]{Wanajo:2013}
{Wanajo}, S. 2013, \apjl, 770, L22, \dodoi{10.1088/2041-8205/770/2/L22}

\bibitem[{{Wanajo} {et~al.}(2018){Wanajo}, {M{\"u}ller}, {Janka}, \&
  {Heger}}]{Wanajo:2018}
{Wanajo}, S., {M{\"u}ller}, B., {Janka}, H.-T., \& {Heger}, A. 2018, \apj, 852,
  40, \dodoi{10.3847/1538-4357/aa9d97}

\bibitem[{{Wanajo} {et~al.}(2009){Wanajo}, {Nomoto}, {Janka}, {Kitaura}, \&
  {M{\"u}ller}}]{Wanajo:2009}
{Wanajo}, S., {Nomoto}, K., {Janka}, H.~T., {Kitaura}, F.~S., \& {M{\"u}ller},
  B. 2009, \apj, 695, 208, \dodoi{10.1088/0004-637X/695/1/208}

\bibitem[{{Wenger} {et~al.}(2000){Wenger}, {Ochsenbein}, {Egret}, {Dubois},
  {Bonnarel}, {Borde}, {Genova}, {Jasniewicz}, {Lalo{\"e}}, {Lesteven}, \&
  {Monier}}]{Simbad}
{Wenger}, M., {Ochsenbein}, F., {Egret}, D., {et~al.} 2000, \aaps, 143, 9,
  \dodoi{10.1051/aas:2000332}

\bibitem[{{Woosley} {et~al.}(2002){Woosley}, {Heger}, \&
  {Weaver}}]{Woosley:2002}
{Woosley}, S.~E., {Heger}, A., \& {Weaver}, T.~A. 2002, Reviews of Modern
  Physics, 74, 1015, \dodoi{10.1103/RevModPhys.74.1015}

\bibitem[{{Yong} {et~al.}(2009){Yong}, {Grundahl}, {D'Antona}, {Karakas},
  {Lattanzio}, \& {Norris}}]{Yong:2009}
{Yong}, D., {Grundahl}, F., {D'Antona}, F., {et~al.} 2009, \apjl, 695, L62,
  \dodoi{10.1088/0004-637X/695/1/L62}

\bibitem[{{Yong} {et~al.}(2017){Yong}, {Norris}, {Da Costa}, {Stanford},
  {Karakas}, {Shingles}, {Hirschi}, \& {Pignatari}}]{Yong:2017}
{Yong}, D., {Norris}, J.~E., {Da Costa}, G.~S., {et~al.} 2017, \apj, 837, 176,
  \dodoi{10.3847/1538-4357/aa6250}

\bibitem[{{Yong} {et~al.}(2014){Yong}, {Roederer}, {Grundahl}, {Da Costa},
  {Karakas}, {Norris}, {Aoki}, {Fishlock}, {Marino}, {Milone}, \&
  {Shingles}}]{Yong:2014}
{Yong}, D., {Roederer}, I.~U., {Grundahl}, F., {et~al.} 2014, \mnras, 441,
  3396, \dodoi{10.1093/mnras/stu806}

\end{thebibliography}
\bibliographystyle{aasjournal}

\end{document}